# THE THEORY OF QUANTITATIVE TRADING

**Andrea Berdondini**

*“The difference between an amateur trader and a professional trader is that the first is obsessed by the result while the second is obsessed by the knowledge”*

# Introduction

The first step that you must take when you want to learn how to develop a quantitative trading system is to answer the following question: what characteristics must a system have to represent the theoretically most difficult situation possible where to make predictions?

The first characteristic that this system must possess is that of having a very low ratio between the deterministic component and the random component. Therefore, the random component must overcome the deterministic component.

The second characteristic that we need to determine concerns the number of degrees of freedom. This parameter determines how much difficult it is to test hypotheses. This consideration derives from the fundamental problem of statistics defined as follows: *"A statistical data does not represent useful information, but becomes useful information only when it is shown that it was not obtained randomly"*. Therefore, given a result, the probability of obtaining it randomly decreases as the degrees of freedom of the system increase. Consequently, systems with a low number of degrees of freedom are particularly dangerous, because it is particularly easy to get good results randomly and therefore you risk overestimating an investment strategy. For this reason, a system that wants to be as difficult as possible where making predictions must have a low number of degrees of freedom.

The third characteristic we have to choose is if to consider the system as ergodic (stationary) or non-ergodic (non-stationary). This choice is easy because it is much more difficult to make forecasts on a non-ergodic system. Indeed, in this case, past results may not be significant with respect to future results.

In conclusion, the system that represents, from the theoretical point of view, the most difficult situation in which to make predictions is a system in which there is a predominant random component, with a low number of degrees of freedom and not ergodic. Is there a system that has all these 3 characteristics? The answer is yes indeed, the financial markets represent a system that respects all these conditions.

I started this book with this consideration because I believe that the most important thing to understand, for any person who wants to develop a quantitative trading system, is to know that you are facing the most difficult situation theoretically possible where to make predictions. It is this awareness that must guide us in the study of the financial markets. Without this awareness, we will inevitably be led to underestimate the difficulty of the problem and this will lead us to make mistakes.

# Structure of the book

This book consists of a selection of articles divided into three main themes:

Statistics

Quantitative Trading

Psychology

These three arguments are indispensable for the development of a quantitative trading system. Although the articles deal with very different topics, they are closely linked to each other, in practice they represent the observation of the same problem from three different points of view.

At the beginning of each chapter there will be an introductory paragraph where the results reported in the articles are summarize. The order of the articles was chosen so as to constitute a single logical reasoning that develops progressively.

# Statistics

Financial markets are characterized by a dominant random component with respect to the deterministic component. For this reason, the articles present are intended to explain the statistical analysis under these conditions.

The main topic that is treated concerns the definition of the uncertainty of the statistical data. The traditional approach considers uncertainty as the dispersion of data around the true value. Therefore, it is based on the hypothesis that any divergence from uniformity is the result of a deterministic component. <u>This condition is realistic only in systems where the random component is negligible. Instead, in cases, such as in finance, where the random component prevails, it turns out to be an unrealistic hypothesis that leads to incorrect conclusions.</u> For this reason, we will give a new definition of uncertainty suitable for a system in which there is a predominant random component. The parameter that has been chosen for its definition is represented by the probability of obtaining an equal or better result in a random way. Knowing how to calculate this parameter correctly represents the basis of statistics in finance. As I will show in the articles, this calculation is very difficult, and extremely easy to underestimate the uncertainty. Indeed, the mistake made is to evaluate the individual hypotheses independently without considering the hypotheses previously tested. This way of operating often leads, in systems with a low number of degrees of freedom where it is easy to obtain good results randomly, to underestimating uncertainty.

This approach will then be applied in different situations, such as in the case of the resolution of the St. Petersburg paradox. In this way, we will have concrete examples of how this method represents a completely new point of view in the evaluation of hypotheses.

# The uncertainty of the statistical data

Andrea Berdondini

ABSTRACT: Any result can be generated randomly and any random result is useless. Traditional methods define uncertainty as a measure of the dispersion around the true value and are based on the hypothesis that any divergence from uniformity is the result of a deterministic event. The problem with this approach is that even non-uniform distributions can be generated randomly and the probability of this event rises as the number of hypotheses tested increases. Consequently, there is a risk of considering a random and therefore non-repeatable hypothesis as deterministic. Indeed, it is believed that this way of acting is the cause of the high number of non-reproducible results. Therefore, we believe that the probability of obtaining an equal or better result randomly is the true uncertainty of the statistical data. Because it represents the probability that the data is useful and therefore the validity of any other analysis depends on this parameter.

## Introduction

Any result can be generated randomly and any random result is useless. Traditional methods [1] and [2] define uncertainty as a measure of the dispersion around the true value and are based on the hypothesis that any divergence from uniformity is the result of a deterministic event. The problem with this approach is that even non-uniform distributions can be generated randomly and the probability of this event rises as the number of hypotheses tested increases. Consequently, there is a risk of considering a random and therefore non-repeatable hypothesis as deterministic. Indeed, it is believed that this way of acting is the cause of the high number of non-reproducible results [3] and [4]. Therefore, we believe that the probability of obtaining an equal or better result randomly is the true uncertainty of the statistical data, because it represents the probability that the data is useful and therefore the validity of any other analysis depends on this parameter.

In addition, we will also address the problem of determining the correct method of calculating the probability of obtaining an equal or better result randomly. Regarding this topic, we will see that the fundamental point, in calculating this probability value, is to consider the statistical data dependent on all the other data generated by all the tested hypotheses.

Considering the statistical data as non-independent has fundamental implications in statistical analysis. Indeed, all our random actions are not only useless, but will increase the uncertainty of the statistical data. For this reason, in the following article [5], we highlight the importance of acting consciously in statistics.

Furthermore, the evaluation of the uncertainty of the statistical data will be possible only by knowing all the attempts made. In practice, the calculation of uncertainty is very difficult because not only we must consider all our attempts, but we must also consider the attempts made by every other person who is performing the same task as us. In this way, the uncertainty of our statistical data also depends on the actions performed by the people who are working our own analysis. Indeed, a group of people who belong to a research network all having the same reputation who all work on the same problem can be considered with one person who carries out all the attempts made. Consequently, the calculation of uncertainty becomes something relative that depends on the information we have.

## Definition of uncertainty

The aim of the definition of uncertainty of the statistical data that we are going to give is to determine a parameter that is linked to the repeatability of the result and that is universal and therefore, independent of the system in which we perform the statistical analysis.

We define the uncertainty of the statistical data as the probability of obtaining an equal or better result randomly.

This definition considers the statistical data as a forecast, so a forecast is repeatable only if the process that generated it is non-random. Consequently, the calculation of uncertainty involves determining the type of process that generated the result. We can distinguish cognitive processes from random processes by their statistical property of generating non-reproducible results in a random way. Indeed, by using the information on the system, on which we are performing a measurement, we can increase our probability of forecasting and this leads to a consequent decrease in the probability of obtaining the same result randomly.

It is interesting to note that the repeatability of the statistical data and non-randomness of the process that produced it are two equivalent concepts. Indeed, the information leads to the repeatability of the result and at the same time generates results that cannot be reproduced randomly.

To understand the definition given, we report the following example: We have to analyze a statistical datum represented by 1000 predictions on an event that can have only two results. The 1000 predictions are divided into 600 successes and 400 failures. To calculate the probability of obtaining an equal or better result in a random way, we use the binomial distribution and we obtain the following value $1.4 \cdot 10^{-8}\%$.

Now, instead, let us consider a statistical datum represented by 10 predictions divided into 8 successes and 2 failures. In this case, the probability of getting an equal or better result randomly is $5.5\%$.

Comparing the two results, we note that in the first case, although the number of successes is only $60\%$, the uncertainty is almost zero, while in the second case, with a probability of success of $80\%$, the uncertainty is much higher. This difference is due to the fact that the definition given, as mentioned, concerns only the repeatability of the result and not its accuracy. Therefore, it is a value that decreases as the repetition of the result increases. The approach presented is very different from the classic approach, where uncertainty is seen as a measure of the dispersion of the data with respect to the true value.

The fundamental point to understand is that the probability that statistical data is completely random and the estimate of its random component (dispersion around the true value) are two parameters that are only partially dependent on each other. The first decreases as the number of repetitions of the measurement increases, the second does not and this is one of the reasons, why the traditional definition of uncertainty, in many cases, is not significant with regard to the repeatability of the result.

The problem, as we have seen in the examples, is that there is always a greater or lesser probability that a purely random process generates the result. In this case, any analysis turns out to be wrong, for this reason, this value is considered the true uncertainty of the statistical result.

## Calculation of the uncertainty of the statistical data

Correctly calculating the probability of getting an equal or better result randomly involves changing our approach to statistics. The approach commonly used in statistics is to consider the data produced by one method independent of the data produced by different methods. This way of proceeding seems the only possible one but, as we will show in the following paradox, it leads to an illogical result, which is instead solved by considering the data as non-independent.

We think to have a computer with enormous computational capacity that is used to develop hypotheses about a phenomenon that we want to study. The computer works as follows: it creates a random hypothesis and then performs a statistical test. At this point, we ask ourselves the following question: can there be a useful statistical test to evaluate the results of the hypothesis generated?

If we answer yes, we get an illogical result because our computer would always be able, by generating a large number of random hypotheses, to find a hypothesis that passes the statistical test. In this way, we arrive at the absurd conclusion that it is possible to create knowledge randomly, because it is enough to have a very powerful computer and a statistical test to understand every phenomenon.

If we answer no, we get another illogical result because we are saying that no hypothesis can be evaluated. In practice, the results of different hypotheses are all equivalent and indistinguishable.

How can we solve this logical paradox? The only way to answer the question, without obtaining an illogical situation, is to consider the results obtained from different methods depending on each other. A function that meets this condition is the probability of getting an equal or better result at random. Indeed, the calculation of this probability implies the random simulation of all the actions performed. Hence, random attempts increase the number of actions performed and consequently increase the probability of obtaining an equal or better result randomly. For this reason, generating random hypotheses is useless, and therefore if you use this parameter, as a measure of uncertainty, it is possible to evaluate the data and at the same time it is impossible to create knowledge by generating random hypotheses.

Considering the statistical data as non-independent is a fundamental condition for correctly calculating the uncertainty. The probability of getting an equal or better result at random meets this condition.

The dependence of statistical data on each other has profound implications in statistics, which will be discussed in the next section.

## Consequences of the non-independence of the statistical data

Considering the statistical data dependent on each other in the calculation of uncertainty leads to three fundamental consequences in statistics.

First fundamental consequence of the non-independence of the statistical data: our every random action always involves an increase in the uncertainty of the statistical data.

Example: We need to analyze a statistical datum represented by 10 predictions about an event that can only have two results. The 10 predictions are divided into 8 successes and 2 failures. To calculate the probability of obtaining an equal or better result randomly we use the binomial distribution and we get the following value 5.5%. If before making these 10 predictions, we tested a different hypothesis with which we made 10 other predictions divided into 5 successes and 5 failures, the uncertainty of our result changes. Indeed, in this case, we must calculate the probability of obtaining a result with a number of successes greater than or equal to 8 by performing two random attempts consisting of 10 predictions each. In this case, the probability becomes 10.6%, so the fact of having first tested a random hypothesis almost doubled the uncertainty of our second hypothesis. Consequently, increasing the random hypotheses increases the number of predictions that we will have to make, with the true hypothesis, to have an acceptable uncertainty.

Second fundamental consequence of the non-independence of the statistical data: every random action of ours and of every other person equivalent to us, always involves an increase in the uncertainty of the statistical data.

By the equivalent term, we mean a person with the same reputation as us, therefore the data produced by equivalent people are judged with the same weight.

Example: 10 people participate in a project whose goal is the development of an algorithm capable of predicting the outcome of an event that can have only two results. An external person who does not participate in the project but is aware of every attempt made by the participants evaluates the statistical data obtained. All participants make 100 predictions, 9 get a 50% chance of success, one gets a 65% chance of success. The uncertainty of the static data of the participant who obtains a probability of success of 65% is obtained by calculating the probability of obtaining a result with a number of successes greater than or equal to 65 by performing ten random attempts consisting of 100 predictions each. The probability obtained, in this way, is 16% instead if he was the only participant in the project the probability would have been 0.18%, therefore about 100 times lower.

Third fundamental consequence of the non-independence of the statistical data: the calculation of the uncertainty varies according to the information possessed.

Example: 10 people participate in a project whose goal is the development of an algorithm capable of predicting the outcome of an event that can have only two results. In this case, people do not know the other participants and think they are the only ones participating in the project. All participants make 100 predictions, 9 get a 50% chance of success and one gets a 65% chance of success. The participant who obtains a probability of success of 65% independently calculates the uncertainty of the result obtained. Not knowing that other people are participating in the project, calculate the probability of obtaining a result with a number of successes greater than or equal to 65 by performing a single random attempt consisting of 100 predictions; the probability obtained is 0.18%. An external person who is aware of every attempt made by the participants calculates the uncertainty of the participant's statistical data, which obtains a probability of success of 65%. It then calculates the probability of obtaining a result with a number of successes greater than or equal to 65 by making ten random attempts consisting of 100 predictions each. The probability obtained, in this way, is 16%, a much higher value than the uncertainty calculated by the participant. The uncertainty value calculated by the external person using more information is most accurate than the uncertainty value calculated by the individual participant. Consequently, the uncertainty value obtained by exploiting the greatest number of information must always be considered, in the case of the example, the most accurate uncertainty is that of 16%.

The first and second fundamental highlighting consequence of the non-independence of the statistical data can be redefined by highlighting the non-randomness of the action.

First fundamental consequence of the non-independence of the statistical data: our every non-random action always involves a decrease in the uncertainty of the statistical data.

Second fundamental consequence of the non-independence of the statistical data: every non-random action of ours and of every other person equivalent to us, always involves a decrease in the uncertainty of the statistical data.

## Conclusion

The traditional definition of uncertainty implies considering true, for non-homogeneous data dispersions, the hypothesis that the result is not completely random. We consider this assumption the main problem of the definition of uncertainty. Indeed, whatever the statistical data obtained, there is always a possibility that they are completely random and therefore useless.

This error stems from the fact that the definition of uncertainty was developed in an environment where each method had a strong deterministic component. Therefore, calculating the probability of obtaining an equal or better result at random might seem useless. However, when we apply statistics in fields such as finance, where the random component is predominant the traditional approach to uncertainty turns out to be unsuccessful. It fails for the simple reason that the hypothesis on which it is based may not be true. For this reason, we have defined the uncertainty of the statistical data as the probability of obtaining an equal or better result randomly. Since this definition of uncertainty is not linked to any hypothesis, it turns out to be universal. The correct calculation of this probability value implies considering the statistical data dependent on each other. This assumption, as we have shown through a paradox, makes the definition of uncertainty given consistent with the logical principle that it is not possible to create knowledge randomly.

The non-independence of the statistical data implies that each action performed has an effect on the calculation of uncertainty. The interesting aspect is that a dependence is also created between actions performed by different people. Consequently, the calculation of uncertainty depends on the information in our possession, so it becomes something relative that can be determined absolutely only with complete knowledge of the information.

## Bibliography:

[1] Bich, W., Cox, M. G., and Harris, P. M. Evolution of the "Guide to the Expression of Uncertainty in Measurement". Metrologia, 43(4):S161–S166, 2006.

[2] Grabe, M .,"Measurement Uncertainties in Science and Technology", Springer 2005.

[3] Munafò, M., Nosek, B., Bishop, D. et al. "A manifesto for reproducible science". Nat Hum Behav 1, 0021 (2017). https://doi.org/10.1038/s41562-016-0021.

[4] Ioannidis, J. P. A. "Why most published research findings are false". PLoS Med. 2, e124 (2005).

[5] Berdondini, Andrea, "Statistics the Science of Awareness" (August 30, 2021). Available at SSRN: https://ssrn.com/abstract=3914134.

# A Modern Approach to the Fundamental Problem of Causal Inference

Andrea Berdondini

ABSTRACT: The fundamental problem of causal inference defines the impossibility of associating a causal link to a correlation, in other words: correlation does not prove causality. This problem can be understood from two points of view: experimental and statistical. The experimental approach tells us that this problem arises from the impossibility of simultaneously observing an event both in the presence and absence of a hypothesis. The statistical approach, on the other hand, suggests that this problem stems from the error of treating tested hypotheses as independent of each other. Modern statistics tends to place greater emphasis on the statistical approach because, compared to the experimental point of view, it also shows us a way to solve the problem. Indeed, when testing many hypotheses, a composite hypothesis is constructed that tends to cover the entire solution space. Consequently, the composite hypothesis can be fitted to any data set by generating a random correlation. Furthermore, the probability that the correlation is random is equal to the probability of obtaining the same result by generating an equivalent number of random hypotheses.

## Introduction

The fundamental problem of causal inference defines the impossibility of associating causality with a correlation; in other words, correlation does not prove causality. This problem can be understood from two perspectives: experimental and statistical. The experimental approach suggests that this problem arises from the impossibility of observing an event both in the presence and absence of a hypothesis simultaneously. The statistical approach, on the other hand, suggests that this problem stems from the error of treating tested hypotheses as independent of each other.

Modern statistics tends to place greater emphasis on the statistical approach, as it, unlike the experimental approach, also provides a path to solving the problem. Indeed, when testing many hypotheses, a composite hypothesis is constructed that tends to cover the entire solution space. Consequently, the composite hypothesis can fit any data series, thereby generating a correlation that does not imply causality.

Furthermore, the probability that the correlation is random is equal to the probability of obtaining the same result by generating an equivalent number of random hypotheses. Regarding this topic, we will see that the key point, in calculating this probability value, is to consider hypotheses as dependent on all other previously tested hypotheses.

Considering the hypothesis as non-independent has fundamental implications in statistical analysis. **In fact, every random action we take is not only useless but will increase the probability of a random correlation.** For this reason, in the following article [1], we highlight the importance of acting consciously in statistics.

Moreover, calculating the probability that the correlation is random is only possible if all prior attempts are known. In practice, calculating this probability is very difficult because not only do we need to consider all our attempts, but we also need to consider the attempts made by everyone else performing the same task. In fact, a group of people belonging to a research network all having the same reputation and all working on the same problem can be considered with a single person who performs all the attempts made. From a practical point of view, we are almost always in the situation where this parameter is underestimated because it is very difficult to know all the hypotheses tested. <u>Consequently, the calculation of the probability that a correlation is casual becomes something relative that depends on the information we have.</u>

## The Fundamental Problem of Causal Inference

The fundamental problem of causal inference [2] defines the impossibility of associating causality with a correlation, in other words: correlation does not prove causality. From a statistical point of view, this indeterminacy arises from the error of considering the tested hypotheses as independent of each other. When a series of hypotheses is generated, a composite hypothesis is formed that tends to fit any data series, leading to purely random correlations.

For example, you can find amusing correlations between very different events on the internet; these correlations are obviously random. These examples are often used to demonstrate the fundamental problem of causal inference**. In presenting this data, the following information is always omitted: how many hypotheses did I consider before finding a related hypothesis.**

This is essential information because if I have a database comprising a very high number of events, for any data series, there will always be a hypothesis that correlates well with my data. Thus, if I generate a large number of random hypotheses, I will almost certainly find a hypothesis that correlates with the data I am studying. Therefore, having a probability of about 100% of being able to obtain the same result randomly, I have a probability of about 100% that the correlation does not also imply causation.

On the other hand, if we generate a single hypothesis that correlates well with the data, in this situation, almost certainly, the correlation also implies causation. This is because the probability of obtaining a good correlation by generating a single random hypothesis is almost zero.

This result is also intuitive, because it is possible to achieve a good correlation with a single attempt only if one has knowledge of the process that generated the data to be analyzed. And it is precisely this knowledge that also determines a causal constraint.

Figure 1 summarizes the basic concepts showing how the correct way to proceed is to consider the hypotheses as non-independent.

The Fundamental Problem of Causal Inference

Correlation does not prove causality

**Wrong way**
Hypotheses are considered independent

A series of hypotheses is generated

Hypotheses 1

Hypotheses 2

.........

Hypotheses n

Hypothesis n is found to be correlated and is analyzed without taking into account the previous hypotheses

**Correct way**
Hypotheses are considered non-independent

A series of hypotheses is generated

Hypotheses 1

Hypotheses 2

A composite hypotheses is formed

When the hypothesis n is found to be correlated with the data; we know that the real correlation is with the composite hypothesis

If you follow the correct method, you will be aware that by generating many hypotheses you are creating a composite hypothesis that will be able to adapt to each series of events, generating a random correlation

*Figure 1: shows that the correct way is to consider the generated hypotheses as non-independent. In this way, one develops the awareness that by generating many hypotheses one creates a composite hypothesis capable of generating random correlations.*

## Calculating the probability that the correlation is random

Correctly calculating the probability of getting an equal or better result randomly involves changing our approach to statistics. The approach commonly used in statistics is to consider the data produced by one method independent of the data produced by different methods. This way of proceeding seems the only possible one but, as we will show in the following paradox, it leads to an illogical result, which is instead solved by considering the data as non-independent.

We think to have a computer with enormous computational capacity that is used to develop hypotheses about a phenomenon that we want to study. The computer works as follows: it creates a random hypothesis and then performs a statistical test. At this point, we ask ourselves the following question: can there be a useful statistical test to evaluate the results of the hypothesis generated?

If we answer yes, we get an illogical result because our computer would always be able, by generating a large number of random hypotheses, to find a hypothesis that passes the statistical test. In this way, we arrive at the absurd conclusion that it is possible to create knowledge randomly, because it is enough to have a very powerful computer and a statistical test to understand every phenomenon.

If we answer no, we get another illogical result because we are saying that no hypothesis can be evaluated. In practice, the results of different hypotheses are all equivalent and indistinguishable.

How can we solve this logical paradox? The only way to answer the question, without obtaining an illogical situation, is to consider the results obtained from different methods depending on each other. A function that meets this condition is the probability of getting an equal or better result at random. Indeed, the calculation of this probability implies the random simulation of all the actions performed. Hence, random attempts increase the number of actions performed and consequently increase the probability of obtaining an equal or better result randomly.

For this reason, generating random hypotheses is useless, and therefore if you use this parameter, it is possible to evaluate the data and at the same time it is impossible to create knowledge by generating random hypotheses. Considering the hypothesis as non-independent is a fundamental condition for correctly calculating of the probability that the correlation is random. The probability of getting an equal or better result at random meets this condition.

The dependence of hypothesis on each other has profound implications in statistics, which will be discussed in the next section.

## Consequences of the non-independence of the hypothesis

Consider the tested hypotheses to be dependent on each other when calculating the probability that the correlation is causal leads to three fundamental consequences in statistics.

First fundamental consequence of the non-independence of the hypothesis: our every random action always involves an increase in the probability of a random correlation.

Example: We need to analyze a statistical datum represented by 10 predictions about an event that can only have two results. The 10 predictions are divided into 8 successes and 2 failures. To calculate the probability of obtaining an equal or better result randomly we use the binomial distribution and we get the following value 5.5%. If before making these 10 predictions, we tested a different hypothesis with which we made 10 other predictions divided into 5 successes and 5 failures, the uncertainty of our result changes. Indeed, in this case, we must calculate the probability of obtaining a result with a number of successes greater than or equal to 8 by performing two random attempts consisting of 10 predictions each. In this case, the probability becomes 10.6%, so the fact of having first tested a random hypothesis almost doubled the probability of a random correlation of our second hypothesis. Consequently, increasing the random hypotheses increases the number of predictions that we will have to make, with the true hypothesis, to have a low probability that the correlation is coincidental.

Second fundamental consequence of the non-independence of the hypothesis: every random action of ours and of every other person equivalent to us, always involves an increase of the probability that the correlation is random.

By the equivalent term, we mean a person with the same reputation as us, therefore the data produced by equivalent people are judged with the same weight.

Example: 10 people participate in a project whose goal is the development of an algorithm capable of predicting the outcome of an event that can have only two results. An external person who does not participate in the project but is aware of every attempt made by the participants evaluates the statistical data obtained. All participants make 100 predictions, 9 get a 50% chance of success, one gets a 65% chance of success. The probability that a 65% success is due to a random correlation is obtained by calculating the probability of obtaining a result with a number of successes greater than or equal to 65 by performing ten random attempts consisting of 100 predictions each. The probability obtained, in this way, is 16% instead if he was the only participant in the project the probability would have been 0.18%, therefore about 100 times lower.

Third fundamental consequence of the non-independence of the hypothesis: the calculation the probability that the correlation is random varies according to the information possessed.

Example: 10 people participate in a project whose goal is the development of an algorithm capable of predicting the outcome of an event that can have only two results. In this case, people do not know the other participants and think they are the only ones participating in the project. All participants make 100 predictions, 9 get a 50% chance of success and one gets a 65% chance of success. The participant who obtains a probability of success of 65% independently calculate the probability that the correlation is coincidental. Not knowing that other people are participating in the project, calculate the probability of obtaining a result with a number of successes greater than or equal to 65 by performing a single random attempt consisting of 100 predictions; the probability obtained is 0.18%. An external person who is aware of every attempt made by the participants calculate the probability that the 65% success rate of one of the participants was due to a random correlation. knowing the number of participants in the project calculates the probability of obtaining a result with a number of successes greater than or equal to 65 by making ten random attempts consisting of 100 predictions each. The probability obtained, in this way, is 16%, a much higher value than the probability calculated by the participant. The probability calculated by the external person using more information is most accurate than the probability calculated by the individual participant. Consequently, the probability obtained by exploiting the greatest number of information must always be considered, in the case of the example, the probability that the 65% success is due to a random correlation is 16%. Therefore, the participant having less information underestimates this probability.

The first and second fundamental highlighting consequence of the non-independence of the hypothesis can be redefined by highlighting the non-randomness of the action.

<u>First fundamental consequence of the non-independence of the hypothesis: our every non-random action always involves a reduction in the probability that the correlation is random.</u>

<u>Second fundamental consequence of the non-independence of the hypothesis: every non-random action of ours and of every other person equivalent to us, always involves a reduction in the probability that the correlation is random.</u>

## How to perform correctly the statistical hypothesis test

About to perform correctly the statistical hypothesis test, It is interesting to note how the non-independence of the hypothesis can be seen as something extremely obvious or as something extremely innovative. Indeed, it may seem absolutely banal to consider all the hypotheses that have been tested, for the obvious reason that by running a large number of random hypotheses sooner or later there will be some hypothesis that will fit the data quite well. On the other hand, also considering the previous hypotheses represents a revolution in the evaluation of a hypothesis. In fact, from this point of view, the mere knowledge of the hypothesis that makes the prediction does not allow us to define its real complexity. Therefore, if in the statistical hypothesis test the p-value [3], [4], used as a threshold to reject the null hypothesis, is calculated considering only the hypothesis that actively participates in the prediction, it means, that we are underestimating the complexity of the hypothesis. Consequently, the p-value, thus calculated, is wrong and therefore determines a false evaluation of the hypothesis. It is therefore believed that this systematic error, in the execution of the hypothesis test, is responsible for the high number of non-reproducible results [5], [6].

Taking advantage of these considerations it is understood that evaluating a statistical result can be very difficult because some information can be hidden. For example, we are obliged to report the mathematical formula that makes the prediction but, instead, we may not report all previous failed attempts. Unfortunately, this information is essential for evaluating the hypothesis, because they are an integral part of the hypothesis. Indeed, if we test 10 hypotheses, we simply interpolate the data with those ten hypotheses and choose the hypothesis that passes the chosen evaluation test. **This problem also depends on the increasing use of statistical software capable of quickly executing** **a huge number of mathematical models.** Consequently, there is the risk of "playing" with this software by performing a multitude of analyzes and this sooner or later leads to a random correlation. For these reasons, the evaluation of statistical results represents one of the most important challenges for scientific research.

Unfortunately, it is a difficult problem to solve because, as mentioned, some information can always be hidden when writing an article. The simplest solution adopted is to use more selective evaluation parameters, which in practice means making it unlikely to pass the evaluation test by developing random hypotheses. However, this solution has a big problem: by acting in this way there is the risk of discarding a correct hypotheses and cannot be applied to all fields of research. For example, in finance where the possible inefficiencies of the markets [7], which can be observed, are minimal, adopting very restrictive valuation methods means having to discard almost any hypothesis.

## Conclusion

In this article, we analyzed the fundamental problem of causal inference from a statistical perspective. From this point of view, the problem arises from treating all tested hypotheses as independent of each other. This way of acting is wrong because when we generate a series of hypotheses we are building a composite hypothesis that will tend to adapt and therefore give a random correlation to each of our series of data. **It is believed that this incorrect approach is the cause of the problem of non-reproducibility of scientific results.** Moreover, the increase in computational capacity speeds up hypothesis development, inadvertently creating composite hypotheses that can lead to random correlations.

The probability that a correlation is random is obtained by calculating the probability of obtaining an equal or better result randomly. This calculation can be done, correctly, only by knowing all the hypotheses tested, unfortunately this information is very difficult to have.

For this reason, in modern statistics it is considered fundamental to develop the awareness that each of our compulsive and irrational actions, which leads us to develop and test a large quantity of hypotheses, has as a consequence the generation of random correlations that are difficult to detect.

## Bibliography:

[1] Berdondini, Andrea, “Statistics the Science of Awareness” (August 30, 2021). Available at SSRN: https://ssrn.com/abstract=3914134.

[2] Holland, P. W. (1986) - Statistics and Causal Inference. Journal of the American Statistical Association, 81(396), 945-960.

[3] Hung, H.M.J., O'Neill, R.T., Bauer, P., & Kohne, K. (1997). "The behavior of the p-value when the alternative hypothesis is true." Biometrics, 53(1), 11–22.

[4] Harlow, L.L., Mulaik, S.A., & Steiger, J.H. (1997). "What if there were no significance tests?" Psychological Methods, 2(4), 315-328.

[5] Munafò, M., Nosek, B., Bishop, D. et al. “A manifesto for reproducible science”. Nat Hum Behav 1, 0021 (2017). https://doi.org/10.1038/s41562-016-0021.

[6] Ioannidis, J. P. A. “Why most published research findings are false”. PLoS Med. 2, e124 (2005).

[7] Black, F. (1971) “Random Walk and Portfolio Management,” Financial Analyst Journal, 27, 16-2

# The information paradox

Andrea Berdondini

ABSTRACT: The following paradox is based on the consideration that the value of a statistical datum does not represent useful information but becomes useful information only when it is possible to prove that it was not obtained in a random way. In practice, the probability of obtaining the same result randomly must be very low in order to consider the result useful. It follows that the value of a statistical datum is something absolute, but its evaluation in order to understand whether it is useful or not is something relative depending on the actions that have been performed. Consequently, a situation such as the one described in this paradox can occur, wherein in one case it is practically certain that the statistical datum is useful, instead of in the other case the statistical datum turns out to be completely devoid of value. This paradox wants to bring attention to the importance of the procedure used to extract statistical information. Indeed, the way in which we act affects the probability of obtaining the same result in a random way and consequently on the evaluation of the statistical parameter.

## The information paradox

We have two identical universes, in both universes the same person is present, that we will call John, he must perform the exact same task which is to analyze a database in order to extract useful correlations. As we have said the universes are equal, so the databases are identical and the person who has to do the work is the same. The database that needs to be analyzed consists of a million parameters related to an event to be studied.

In the universe "1", John acts as follows: he takes the whole database and calculates the correlation of the parameters with the event to be studied. From this analysis he finds 50 parameters with a high correlation with the event, the correlation found has a chance to happen randomly of 0.005%. Of these 50 parameters, John identifies 10 that according to his experience can be useful in order to study the event. However it is important to point out that the assumptions made by John, on the 10 parameters, are only hypotheses based on his experience, they are not scientific demonstrations that explain precisely the correlation of the 10 parameters with the event.

In the universe "2", John acts in the following way: before analyzing the entire database he uses his knowledge of the event in order to select 10 parameters, that he believes are most correlated with the event, from the million parameters available. However, also in this case, it is important to point out that the assumptions made by John, on the 10 parameters, are only hypotheses based on his experience, they are not scientific demonstrations that explain precisely the correlation of the 10 parameters with the event. Analyzing only these 10 parameters, he finds 5 of them with a high correlation with the event, the correlation found has a chance to happen randomly of 0.005% (as in the previous case).

In practice, the fundamental difference in the analysis method that John does in the two universes is that: in the first universe John uses his own experience after performing statistical analysis on the whole database, instead in the second universe, John uses his experience before to perform the statistical analysis in order to reduce the size of the database.

Now let us see how this different approach affects the evaluation of the data obtained. To do this, we must calculate the probability of obtaining the same results randomly in the two cases.

In the first case, universe "1", in order to calculate the probability of obtaining the same results in a random way we must use the binomial distribution formula with the following parameters:

probability of victory (p) = probability of getting the same correlation randomly

number of successes (k) = number of parameters that present the correlation considered

number of tests (L) = total number of parameters present in the database

By entering these data within the binomial distribution formula:

$$F(k, L, p) = \binom{L}{k} p^k (1-p)^{L-k}$$

p = 0.005%

k = 50

L = 1 Million

We get a probability of 5.6% as a result.

Now let's consider the second case, the universe "2", even in this situation, in order to calculate the probability of obtaining the same results in a random way we must use the binomial distribution formula with the following parameters:

p = 0.005%

k = 5

L = 10

The probability obtained in this case is $7.9 \cdot 10^{-18}$ %.

Analyzing these results it is easy to understand that a percentage of 5.6% makes the correlations found not significant. In order to understand how high this percentage is, we can also calculate the probability of obtaining, in a random way, more than 50 of parameters with the correlation considered, this probability is 46%.

Now we analyze the percentage of the second case ($7.9 \cdot 10^{-18}$ %) this percentage is extremely low, consequently we are practically certain that the correlation found is not random and therefore this result represents a useful information for studying the event.

At this point, John must to decide whether to implement the correlations found or not. Obviously, exploiting the correlations found implies costs, therefore a wrong evaluation involves a high risk. In the universe “1” John is in a difficult situation, in fact the work done is not only useless but also dangerous because it can lead him to sustain wrong investments. Instead, in the second universe John knows that the probability that the correlation is random is almost zero, so he can invest with an acceptable risk.

In conclusion, a simple procedural error has led to enormous consequences. In the first case the experience of john is useless, instead in the second case it was a key resource in order to extract useful information from a big database.

In fact, in the case of the universe “1”, John can no longer use his own knowledge and the only thing he can do is transform his hypotheses into real scientific demonstrations, but in many situations, as in the financial field, doing it can be very difficult. Consequently, when hypotheses are made after having carried out an analysis, these hypotheses risk being conditioned by the results and therefore lose value. Instead, the hypotheses made before the analysis are not conditioned and the analysis of the data is used in order to verify them in a statistical way, as happened in the universe “2”.

One of the fields where it is fundamental to calculate the probability of obtaining the same data in a random way, as a method of evaluating the correlations detected, is the financial one [1], [2].

## Conclusion

In this article we have used a paradox to explain how a statistical datum does not represent a useful information, it becomes a useful information, to study an event, only when it is possible to prove that the probability that it was obtained in a random way is very low. This consideration makes the application of statistics, as a method of evaluating a hypothesis, a "relativistic" science. In fact, as described in the paradox, the calculation of the probability of obtaining the same result in a random way is something of relative that depend from the method used and from the actions performed.

These considerations have a great impact from an experimental point of view, because they teach us the importance of correct planning, in which we must always implement all the knowledge about the event we want to study. It is also essential keep track of all the operations performed on the data, because this information is necessary in order to calculate correctly the probability of obtaining the same results in a random way.

This way of interpreting statistical data is also very useful for understanding the phenomenon of overfitting, a very important issue in data analysis [3], [4]. The overfitting seen from this point of view is simply the direct consequence of considering the statistical parameters, and therefore the results obtained, as a useful information without checking  that them was not obtained in a random way. Therefore, in order to estimate the presence of overfitting we have to use the algorithm on a database equivalent to the real one but with randomly generated values, repeating this operation many times we can estimate the probability of obtaining equal or better results in a random way. If this probability is high, we are most likely in an overfitting situation. For example, the probability that a fourth-degree polynomial has a correlation of 1 with 5 random points on a plane is 100%, so this correlation is useless and we are in an overfitting situation.

This approach can also be applied to the St Petersburg paradox [5], in fact also in this case the expectation gain is a statistical datum that must be evaluated before being used at the decisional level. In fact, the difficulty in solving the paradox stems from the fact of considering a statistical datum always as a useful information. Analyzing the expectation gain it is possible to proof that we can obtain better result, randomly, with a probability that tends asymptotically to 50%. Consequently, the expectation gain that tends to infinity turns out to be a statistic data without value that cannot be used for decision-making purposes.

This way of thinking gives an explanation to the logical principle of Occam's razor, in which it is advisable to choose the simplest solution among the available solutions. In fact, for example, if we want to analyze some points on a plane with a polynomial, increasing the degree increases the probability that a given correlation can occur randomly. For example, given 24 points on a plane, a second degree polynomial has a 50% probability of randomly having a correlation greater than 0.27, instead a fourth degree polynomial has a probability of 84% of having a correlation greater than 0.27 randomly. Therefore, the value of the correlation is an absolute datum but its validity to study a set of data is something relative that depends on the method used. Consequently the simpler methods, being less parameterized, have a lower probability of a randomly correlation, so they are preferred over the complex methods.

## References

[1] Andrea Berdondini, "Application of the Von Mises' Axiom of Randomness on the Forecasts Concerning the Dynamics of a Non-Stationary System Described by a Numerical Sequence" (January 21, 2019). Available at SSRN: https://ssrn.com/abstract=3319864 or http://dx.doi.org/10.2139/ssrn.3319864.

[2] Andrea Berdondini, "Description of a Methodology from Econophysics as a Verification Technique for a Financial Strategy", (May 1, 2017). Available at SSRN: https://ssrn.com/abstract=3184781.

[3] Igor V. Tetko, David J. Livingstone, and Alexander I. Luik, "Neural network studies. 1. Comparison of overfitting and overtraining", Journal of Chemical Information and Computer Sciences 1995 35 (5), 826-833 DOI: 10.1021/ci00027a006.

[4] Quinlan, J.R. (1986). "The effect of noise on concept learning". In R.S. Michalski, J.G. Carbonell, & T.M. Mitchell (Eds.),Machine learning: An artificial intelligence approach(Vol. 2). San Mateo, CA: Morgan Kaufmann.

[5] Andrea Berdondini, "Resolution of the St. Petersburg Paradox Using Von Mises' Axiom of Randomness" (June 3, 2019). Available at SSRN: https://ssrn.com/abstract=3398208.

# Use of the fundamental problem of statistics to define the validity limit of Occam's razor principle

Andrea Berdondini

ABSTRACT: In statistics, to evaluate the significance of a result, one of the most used methods is the statistical hypothesis test. Using this theory, the fundamental problem of statistics can be expressed as follows: "*A statistical data does not represent useful information, but becomes useful information only when it is shown that it was not obtained randomly*". Consequently, according to this point of view, among the hypotheses that perform the same prediction, we must choose the result that has a lower probability of being produced randomly. Therefore, the fundamental aspect of this approach is to calculate correctly this probability value. This problem is addressed by redefining what is meant by hypothesis. The traditional approach considers the hypothesis as the set of rules that actively participate in the forecast. Instead, we consider as hypotheses the sum of all the hypotheses made, also considering the hypotheses preceding the one used. Therefore, each time a prediction is made, our hypothesis increases in complexity and consequently increases its ability to adapt to a random data set. In this way, the complexity of a hypothesis can be precisely determined only if all previous attempts are known. Consequently, Occam's razor principle no longer has a general value, but its application depends on the information we have on the tested hypotheses.

## Introduction

The logical principle of Occam's razor [1], [2], suggests choosing the simplest hypothesis among those available. In this article, we will analyze this principle using the theory of statistical hypothesis test [3], [4]. By exploiting this theory, we will reformulate the fundamental problem of statistics in such a way as to bring attention to the link between the statistical data and the probability that it was produced randomly. Consequently, according to this point of view, among the hypotheses that perform the same prediction, we must choose the result that has a lower probability of being produced randomly. Therefore, it becomes essential to calculate this probability value correctly.

This problem is addressed by redefining what is meant by hypothesis. The traditional approach considers the hypothesis as the set of rules that actively participate in the forecast. Instead, we consider as hypotheses the sum of all the hypotheses made, also considering the hypotheses preceding the one used. Therefore, each time a prediction is made, our hypothesis increases in complexity and consequently increases its ability to adapt to a random data set. In this way, the complexity of a hypothesis can be precisely determined only if all previous attempts are known. Consequently, Occam's razor principle no longer has a general value, but its application depends on the information we have on the tested hypotheses.

Finally, we use this new definition of hypothesis to understand the reason for the high percentage of non-reproducible results, in which the hypothesis test was used.

## The fundamental problem of statistics

In statistics, to evaluate the significance of a result, one of the most used methods is the statistical hypothesis test. Using this theory, the fundamental problem of statistics can be expressed as follows: "*A statistical data does not represent useful information, but becomes useful information only when it is shown that it was not obtained randomly*".

This definition is particularly significant, because it highlights the two fundamental aspects of statistics, which are its uncertainty and the reason for its uncertainty. Indeed, the purpose of statistics is the study of phenomena in conditions of uncertainty or non-determinism by exploiting the sampling of events related to the phenomenon to be studied. Knowing that the observed events can be randomly reproduced with a probability that will never be zero, we understand the reason for the indeterminism that characterizes the statistics. This probability value is called universal probability [5].

Through this definition of the fundamental problem of statistics, it is also possible to formulate the following paradox [6], which highlights how the evaluation of statistical results is dependent on each action performed on the analyzed data.

## The validity limit of Occam's razor principle

In this paragraph, we will see how the information regarding the development of a hypothesis is fundamental to define the validity limit of Occam's razor principle.

Let us start by giving some definitions useful to formalize our theory.

Given an experiment that measures *N* values of a discrete variable *X* with cardinality $C$, we call $D$ the set of dimension $C^N$, which includes all possible sequences $X^N$ of length *N* that can be observed.

Now, we redefine the concept of hypothesis in order to define a chronological succession among the tested hypothesis.

We call *H(t)* the hypothesis developed at time *t*.

We call *PH(t)* the set of sequences $X^N \in D$ that the hypothesis *H(t)* is able to predict.

We call *NPH(t)* the cardinality of the set *PH(t)*.

We call *TH(t)* the set that includes all the hypotheses up to time *t*.

$$TH(t) = \{H(i_1), H(i_2), \dots \dots, H(i_t)\}$$

We call *TPH(t)* the union of all the sets *PH(t)* relating to all the hypotheses *H(t)* ∈ *TH(t)*.

$$TPH(t) = \bigcup_{i=0}^{t} PH(i)$$

We call *NTPH(t)* the cardinality of the set *TPH(t)*. Consequently, *NTPH(T)* defines the number of sequences, belonging to *D,* that the hypothesis *TH(t)* is able to predict. It may happen that different hypotheses forecast the same sequence of values of *X*, having made the union of the sets *PH(t)* these sequences are calculated only once.

If we have only made a hypothesis *H(t)=TH(t)* and *NPH(t)=NTPH(t)*.

If, on the other hand, more than one hypothesis has been tested $H(t) \neq TH(t)$ and $NPH(t) \leq NTPH(t)$.

We define the ability of the hypothesis $TH(t)$ to predict a sequence of $N$ casual observations of the i.i.d. random variable $X$ with discrete uniform distribution and cardinality $C$, the ratio:

$$\frac{NTPH(t)}{C^N} \tag{1}$$

This ratio also defines the probability that the hypothesis $TH(t)$ can predict the results of an experiment, in which the cardinality of $D$ is equal to $C^N$, in a completely random way.

Knowing that a hypothesis $TH(t)$ can predict the results of an experiment only in the following two conditions:

1) $TH(t)$ is true.

2) $TH(t)$ is false and the prediction occurs randomly; the probability of this event is given by equation (1).

Under these conditions, the probability that the hypothesis $TH(t)$ is true turns out to be:

$$1 - \frac{NTPH(t)}{C^N} \tag{2}$$

Consequently, this equation defines the parameter that must be used in the evaluation of $H(t)$. So, if we want to compare two hypotheses $H1(t)$ and $H2(t)$, we have 4 possible results:

1) $NPH1(t) > NPH2(t)$ and $NTPH1(t) > NTPH2(t)$

2) $NPH1(t) > NPH2(t)$ and $NTPH1(t) < NTPH2(t)$

3) $NPH1(t) < NPH2(t)$ and $NTPH1(t) < NTPH2(t)$

4) $NPH1(t) < NPH2(t)$ and $NTPH1(t) > NTPH2(t)$

$NPH(t)$ and $NTPH(t)$ define the number of sequences that hypothesis $H(t)$ and the hypothesis $TH(t)$ are able to predict. Consequently, they can be used as a measure of their complexity, in fact, the more complex a hypothesis is, the greater the number of results it can predict.

Analyzing the four possible results, we note that even if a hypothesis $H1(t)$ is less complex than a hypothesis $H2(t)$ ($NPH1(t) < NPH2(t)$), it is possible to have a hypothesis $TH1(t)$ more complex than a hypothesis $TH2(t)$ ($NTPH1(t) > NTPH2(t)$). Consequently, using equation (2) as an evaluation method, hypothesis $H1(t)$ should be discarded in favor of hypothesis $H2(t)$. This situation can happen, for example, if $H1(t)$ is the last hypothesis of a long series of other hypotheses tested previously.

In the event that there is no information on the hypotheses to be evaluated, it must be assumed that the hypotheses have been developed under the same conditions. Therefore, in this case, not being able to calculate $TH(t)$, it is recommended to choose the simpler hypothesis $H(t)$.

Finally, from equation (2), we can deduce the following result: given a hypothesis $H(t)$ the probability that is true can be calculated only if all the previously tested hypotheses are known.

Consequently, the complexity of a hypothesis not only depends on the mathematical formula that makes the prediction, but also depends on all the attempts made previously. Therefore, Occam's razor principle does not have an absolute value but its application depends on the information about the hypotheses.

## How to perform correctly the statistical hypothesis test

It is interesting to note how the definition of hypothesis, which was given in the previous paragraph, can be seen as something extremely obvious or as something extremely innovative. Indeed, it may seem absolutely banal to consider all the hypotheses that have been tested, for the obvious reason that by running a large number of random hypotheses sooner or later there will be some hypothesis that will fit the data quite well. On the other hand, also considering the previous hypotheses represents a revolution in the evaluation of a hypothesis. In fact, from this point of view, the mere knowledge of the hypothesis that makes the prediction does not allow us to define its real complexity.

Therefore, if in the statistical hypothesis test the p-value [7], [8], used as a threshold to reject the null hypothesis, is calculated considering only the hypothesis that actively participates in the prediction, it means, that we are underestimating the complexity of the hypothesis. Consequently, the p-value, thus calculated, is wrong and therefore determines a false evaluation of the hypothesis. It is therefore believed that this systematic error, in the execution of the hypothesis test, is responsible for the high number of non-reproducible results [9], [10].

Taking advantage of these considerations it is understood that evaluating a statistical result can be very difficult because some information can be hidden. For example, we are obliged to report the mathematical formula that makes the prediction but, instead, we may not report all previous failed attempts. Unfortunately, this information is essential for evaluating the hypothesis, because they are an integral part of the hypothesis. Indeed, if we test 10 hypotheses, we simply interpolate the data with those ten hypotheses and choose the hypothesis that passes the chosen evaluation test.

This problem also depends on the increasing use of statistical software capable of quickly executing a huge number of mathematical models. Consequently, there is the risk of "playing" with this software by performing a multitude of analyzes and this sooner or later leads to a random correlation.

For these reasons, the evaluation of statistical results represents one of the most important challenges for scientific research. Unfortunately, it is a difficult problem to solve because, as mentioned, some information can always be hidden when writing an article. The simplest solution adopted is to use more selective evaluation parameters, which in practice means making it unlikely to pass the evaluation test by developing random hypotheses. However, this solution has different problems in fact, in this way, we can discard correct hypotheses and cannot be applied to all fields of research. For example, in finance where the possible inefficiencies of the markets [11], which can be observed, are minimal, adopting very restrictive valuation methods means having to discard almost any hypothesis.

## Conclusion

In this article, we have discussed the logical principle of Occam's razor using the hypothesis test theory. This allowed us to reformulate the fundamental problem of statistics, in such a way as to make us understand the importance of correctly calculating the probability of obtaining the same results randomly. Solving this problem involved redefining the concept of hypothesis. According to this point of view, by hypothesis we mean the sum of all tested hypotheses. Consequently, the complexity of a hypothesis not only depends on the mathematical formula that makes the prediction but also depends on the previous hypotheses tested.

Therefore, according to this approach, the logical principle of Occam's razor no longer has a general value if one considers as a hypothesis only the set of rules that actively participate in the prediction. If, on the other hand, the hypothesis is considered as the sum of all the tested hypotheses, in this case, Occam's razor principle returns to have a general value.

Finally, it is noted that not considering all the tested hypotheses causes a systematic error in the application of the statistical hypothesis test. Therefore, it is hypothesized that this error, which leads to underestimate the complexity of a hypothesis, is the cause of the high percentage of non-reproducible scientific results.

## References

[1] Roger Ariew, “Ockham's Razor: A Historical and Philosophical Analysis of Ockham's Principle of Parsimony”, 1976.

[2] Sober, Elliott (2004). "What is the Problem of Simplicity?". In Zellner, Arnold; Keuzenkamp, Hugo A.; McAleer, Michael (eds.). Simplicity, Inference and Modeling: Keeping it Sophisticatedly Simple. Cambridge, U.K.: Cambridge University Press. pp. 13–31. ISBN 978-0-521-80361-8. Retrieved 4 August 2012ISBN 0-511-00748-5 (eBook [Adobe Reader]) paper as pdf.

[3] Fisher, R (1955). "Statistical Methods and Scientific Induction" (PDF). Journal of the Royal Statistical Society, Series B. 17 (1): 69–78.

[4] Borror, Connie M. (2009). "Statistical decision making". The Certified Quality Engineer Handbook (3rd ed.). Milwaukee, WI: ASQ Quality Press. pp. 418–472. ISBN 978-0-873-89745-7.

[5] Cristian S. Calude (2002). “Information and Randomness: An Algorithmic Perspective”, second edition. Springer. ISBN 3-540-43466-6.

[6] Berdondini Andrea, “The Information Paradox”, (July 8, 2019). Available at SSRN: https://ssrn.com/abstract=3416559.

[7] Wasserstein, Ronald L.; Lazar, Nicole A. (7 March 2016). "The ASA's Statement on p-Values: Context, Process, and Purpose". The American Statistician. 70 (2): 129–133. doi:10.1080/00031305.2016.1154108.

[8] Hung, H.M.J.; O'Neill, R.T.; Bauer, P.; Kohne, K. (1997). "The behavior of the p-value when the alternative hypothesis is true". Biometrics (Submitted manuscript). 53 (1): 11–22.

[9] Munafò, M., Nosek, B., Bishop, D. et al. “A manifesto for reproducible science”. Nat Hum Behav 1, 0021 (2017). https://doi.org/10.1038/s41562-016-0021.

[10] Ioannidis, J. P. A. “Why most published research findings are false”. PLoS Med. 2, e124 (2005).

[11] Black, F. (1971) “Random Walk and Portfolio Management,” Financial Analyst Journal, 27, 16-22.

# Resolution of the St. Petersburg paradox using Von Mises' axiom of randomness

Andrea Berdondini

ABSTRACT: In this article we will propose a completely new point of view for solving one of the most important paradoxes concerning game theory. The method used derives from the study of non-ergodic systems. This circumstance may create a dependency between results that are often extremely difficult to detect and quantify, such as in the field of finance. Consequently, the expected gain obtained from data that may be correlated has a statistical value that is difficult to determine, thus it cannot be used for decision-making purposes. Therefore, in this scenario, an alternative parameter to be use during the decision-making process must be found. The solution develop shifts the focus from the result to the strategy's ability to operate in a cognitive way by exploiting useful information about the system. In order to determine from a mathematical point of view if a strategy is cognitive, we use Von Mises' axiom of randomness. Based on this axiom, the knowledge of useful information consequently generates results that cannot be reproduced randomly. Useful information in this case may be seen as a significant datum for the recipient, for their present or future decision-making process. In conclusion, the infinite behaviour in this paradox may be seen as an element capable of rendering the expected gain unusable for decision-making purposes. As a result, we are forced to face the problem by employing a different point of view. In order to do this we shift the focus from the result to the strategy's ability to operate in a cognitive way by exploiting useful information about the system. Finally, by resolving the paradox from this new point of view, we will demonstrate that an expected gain that tends toward infinity is not always a consequence of a cognitive and non-random strategy. Therefore, this result leads us to define a hierarchy of values in decision-making, where the cognitive aspect, whose statistical consequence is a divergence from random behaviour, turns out to be more important than the expected gain.

## Introduction

The St. Petersburg paradox represents one of the most important paradoxes in game theory. The classic solution used to solve uses special utility functions that implement the concept of marginal utility [1], [2], [3]. This type of approach has been strongly criticized in virtue of the fact that utility functions attempt to formalize sociological behaviour from a mathematical point of view, and this is why they always have a subjectivity component. Moreover, many studies of behavioural economics [4], [5] highlight how people's behaviour is often irrational. Consequently, the resolution of this paradox still represents an open challenge and, as we will see, the search for an alternative solution may help us improve the decision-making process during the evaluation of a strategy.

In order to understand the method proposed in this article for resolving the St. Petersburg paradox, we must first explain the origins of this method. This approach was developed to study the strategies operating on non-ergodic systems. In particular, the primary field of application is characterized by the study of quantitative trading algorithms operating on financial markets. The non-ergodicity condition can make the results dependent on each other. Therefore, in this scenario a dependency is created between data which, like in the field of finance, is difficult to detect and quantify.

To explain to you the possible consequences at a decision-making level of this condition, I propose the following example: think about making a hundred bets on a hundred flips of a coin and winning one hundred times. In this case, you will have obtained a hundred victories independent of each other to which a very high-expected gain will be associated. You therefore reach the right conclusion that the strategy used to predict the flip of the coin is most likely correct. Now let's change the starting scenario and let's say we make 100 equal bets on a single coin flip, getting a hundred wins. Obviously, since the bets are completely dependent on each other, in this case they cannot be used to calculate the expected gain. In fact, it's basically as if we made a single bet. Now let's imagine, as a third and last scenario, that you are not able to see the person flipping the coin: if we win all the hundred times, we don't know if they are dependent or independent of each other. This third scenario generates a very important decision-making problem, because if I consider the results as independent and they are not, I risk overestimating the strategy. This wrong assessment can lead me to make the irrational choice of using a useless strategy. In finance the non-independence of the results creates a statistical phenomenon called clustering. This statistical characteristic determines the formation of groups of high returns, alternating with groups of low returns. In other words, it means that returns are not distributed evenly but tend to cluster together. The clustering phenomenon has disastrous effects in finance, because when you are going through winning phases you are led to consider the operations carried out as independent of each other. This implies that the expected gain, calculated from data that we mistakenly think to be independent, is overestimated, therefore the evaluation of the strategy will also be incorrect. So, this behaviour can subject us to unexpected risk. With regards to this topic we have developed a paradox [6], which we have called "the professional trader's paradox". The name derives from the fact that we are inclined to consider that our operations are always independent, and therefore, when we face a series of winning bets, we tend to overestimate the strategy used.

In conclusion, the expected gain, obtained from data that may not be independent, cannot be used for decision-making purposes, as it has a statistical value that is difficult to determine. Therefore, from this example we understand that there are situations, like in non-ergodic systems, where the expected gain is no longer a reliable parameter. Consequently, we can think that other situations may exist, like in the case of the infinite behaviour of this paradox, where the expected gain is a datum that cannot be used in decision-making.

We begin to understand that the problem in resolving the St. Petersburg paradox may derive from considering the expected gain, and its variants (utility functions), as the only possible point of view in the evaluation of a strategy. So, the question we have to ask ourselves is: is there a parameter that is better than the expected gain?

The answer we give to this question is focused on being able to understand, from a statistical point of view, if a strategy operates in a cognitive way by exploiting useful information present on the system. The useful information in this case can be seen as a datum subject to analysis that rendered it significant to the recipient for their present or future decision-making process. To determine mathematically if a strategy is cognitive, in the sense just described, we exploit the Von Mises' axiom of randomness. The axiom defines the statistical characteristic that must have a sequence in order to be considered random. The axiom is the following: "the essential requirement for a sequence to be defined as random consists in the complete absence of any rules that may be successfully applied to improve predictions about the next number ".

The meaning of this axiom is the following: when we understand a set of rules to which a numerical sequence is subject we can obtain results, intended as forecasts on the next number of the sequence, whose probability of being reproduced randomly tends toward zero on increasing the number of forecasts made.

Consequently, the results obtained with a game strategy that implements information useful to improve our probability of winning, generates results that cannot be reproduced randomly. Basically, the probability of obtaining better results with a random strategy compared to a cognitive strategy, which implements useful information, tends toward zero as the number of predictions made increases.

This axiom is indeed a statistical method for evaluating the results obtained, without taking into consideration the absolute value of the expected gain. In fact, this method is based solely on the fact of being able to discriminate whether the results were obtained with a random strategy or through a cognitive strategy that implements a set of rules to which the system is subject.

In this way we obtain a fundamental result in analysing the strategies of a particular class of zero-sum games, where there is a balance between the participants. Balance means the situation where none of the players has an implicit advantage over the others. The famous mathematician Daniel Bernoulli defined this particular class of games as: “mathematically fair game of chance”.

This type of game plays a particularly important role in game theory, because it represents a very frequent situation in various fields of interest such as finance.

If we analyse the results obtained by repeating the game of chance described in the St. Petersburg paradox a large number of times, in the next paragraph we will demonstrate that better profits can be obtained with a probability that tends toward 50% by using a random strategy. In practice, the results of a purely random game strategy tend to be distributed symmetrically with respect to the expected theoretical gain derived from the strategy described in the paradox. This result indicates that the doubling-down strategy after each lost bet does not exploit any kind of useful information, and therefore it is a completely non-cognitive game method. Consequently, by taking the cognitive aspect as a parameter to be used in decision-making, and having demonstrated the complete absence within the strategy, we are able to solve the paradox by proving the irrationality of the game method.

In this article, we want to introduce the cognitive aspect, understood in the sense of acting in a non-random way by exploiting useful information about the system, as a fundamental element for improving the decisions theory. In fact, this paradox is useful to make us understand that the knowledge of useful information about the system, capable of increasing our probability of victory, always involves an increase of the expected gain. However, the opposite is not true: an expected gain that tends to infinity does not imply that the strategy exploits knowledge about the system and therefore is cognitive and not random. Consequently, a hierarchy of values is created, where the cognitive aspect is more important than the expected gain for decision-making purposes.

## Resolution of the St. Petersburg paradox

In this paragraph, we will solve the St. Petersburg paradox by demonstrating that the doubling-down strategy after each lost bet is a non-cognitive strategy, which implements no useful information that can be used to improve the probability of success.

In order to do this we have to define the random strategy, which we will use to calculate the probability of obtaining better results than those obtained with the gambling method defined in the paradox. In fact, as mentioned in the previous paragraph, this probability should tend to zero if the strategy being evaluated is a cognitive strategy that implements useful information. Firstly, we define some parameters that are fundamental to characterize our random strategy of reference.

The first parameter we need is the expected value EV of the game obtained by using the doubling-down strategy after each bet lost. Given a number of flips equal to L, with a 50% probability of winning and placing the value of the first bet equal to 1, we have:

$$EV = \frac{L}{2}$$

The second parameter is the average bet AB. By carrying out L bets of Bn value, we have:

$$AB = \frac{(B1 + B2 \ldots + BL)}{L}$$

Knowing that the first bet B1 is equal to 1, and double downing after every bet lost and returning to value 1 when we win the bet, we have:

$$AB = \sum_{n=2}^{L} \frac{(L-n+1)2^{n-1}}{2^n L} + 1$$

$$AB = \sum_{n=2}^{L} \frac{(L-n+1)}{2L} + 1$$

$$AB = \frac{L(L-1)}{2}\frac{1}{2L} + 1$$

$$AB = \frac{L-1}{4} + 1$$

At this point the random game strategy will be defined as follows: given a number of flips equal to L, L bets of AB constant value will be made, randomly choosing whether to bet heads or tails on each bet. To calculate whether a strategy of this type can obtain better results compared to the expected value EV of the strategy of the paradox, just use the binomial distribution formula.

$$F(k, L, p) = \binom{L}{k} p^k (1-p)^{L-k}$$

P = probability of winning

K = number of wins

L = number of tosses

By using the binomial distribution formula, given a value of L, we can obtain the probability of achieving better results with the random strategy described above. The results for the L values ranging from 10 to 200 are shown in Figure 1. Looking at the figure, we see how the probability tends asymptotically toward 50%. Therefore, we have a 50% chance of getting better or worse results. Basically, the strategy described in the paradox tends asymptotically toward a random strategy. Consequently, the doubling-down strategy turns out to be a strategy that does not implement useful information for improving our likelihood of victory. Thus, using the cognitive aspect as a method of evaluation has proven the irrationality of the strategy.

We use very similar approaches, where the strategy being evaluated is compared with an equivalent random strategy, in the financial field to analyse the results generated by a trading strategy [7], [8].

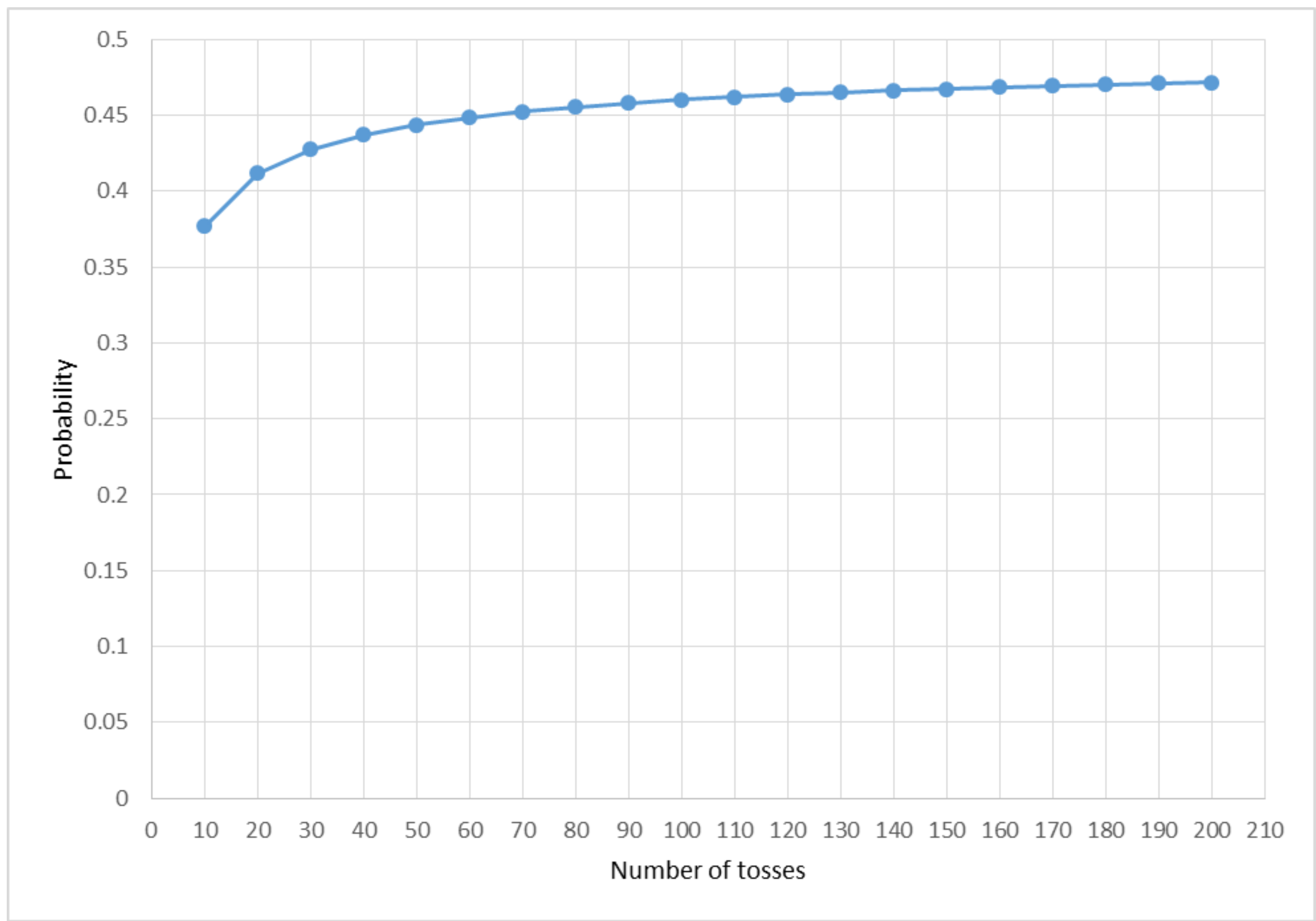

**FIG. 1**: Probability of obtaining better results with a random strategy, given a number of L tosses ranging from 10 to 200.

## Conclusion

In this article we use the St. Petersburg paradox to introduce a parameter related to the cognitive aspect of a strategy, as a fundamental element to help our decision-making in all those situations where the expected gain turns out to be an unreliable parameter. This approach was developed by studying non-ergodic systems. In this scenario the results can be non-independent, so the expected gain becomes a parameter with a statistical value that is difficult to determine. Therefore, it cannot be used in decision-making and a new parameter needs to be found. The parameter chosen is related to the strategy's ability to operate in a cognitive way (the cognitive term indicates the strategy's ability to operate in a non-random way by exploiting useful information about the system, capable of making us increase the probability of victory).

To determine mathematically if a strategy is cognitive, we used the von Mises' axiom of randomness. Based on this axiom, strategies that implement useful information about the system generate results that cannot be reproduced randomly. Thus, we compared the paradox strategy with a completely random but equivalent strategy from the point of view of the total betting value. From this comparison, we have demonstrated that the random strategy gets better results with a probability that tends toward 50% as the number of tosses increases. Basically, the strategy tends to converge to a random strategy instead of diverging as we would expect from a cognitive strategy. In fact, if a strategy implements useful information on the system, the probability of randomly obtaining better results tends toward zero. This result indicates that the doubling-down strategy after each lost bet is not a cognitive strategy that exploits useful information about the system, and therefore by taking the cognitive aspect as an evaluation parameter we have solved the paradox.

In conclusion, the St. Petersburg paradox teaches us that an expected gain that tends toward infinity does not always imply the presence of a cognitive and non-random strategy. Thus knowledge, meaning the exploitation of useful information capable of making us increase the probability of victory, always implies an increase in the expected gain, but the opposite is not true; an expected gain that tends toward infinity can also be obtained in the absence of knowledge about the system. Consequently, from the decision-making aspect we can create a hierarchy of values, where knowledge is more important than the expected gain. In fact, the expectation of victory can be difficult to estimate as in the case of non-ergodic systems or be a non-useful data if the developed strategy has a high degree of overfitting. In all these cases the calculation of the probability of obtaining the same results randomly becomes a much more reliable parameter, since this datum is influenced only by the real knowledge we have of the system and not by the noise. In fact, a statistical datum does not represent a useful information, but becomes a useful information only when it is possible to proof that it was not obtained in a random way. In practice, the probability of obtaining the same result randomly must be very low in order to consider the result useful.

## References

[1] Paul Samuelson, (March 1977). "St. Petersburg Paradoxes: Defanged, Dissected, and Historically Described". Journal of Economic Literature. American Economic Association. 15 (1): 24–55. JSTOR 2722712.

[2] Robert J.Aumann, (April 1977). "The St. Petersburg paradox: A discussion of some recent comments". Journal of Economic Theory. 14 (2): 443–445. doi:10.1016/0022-0531(77)90143-0.

[3] Robert Martin, (2004). "The St. Petersburg Paradox". In Edward N. Zalta. The Stanford Encyclopedia of Philosophy (Fall 2004 ed.). Stanford, California: Stanford University. ISSN 1095-5054. Retrieved 2006-05-30.

[4] Daniel Kahneman and Amos Tversky (1979) "Prospect Theory: An Analysis of Decision under Risk", Econometrica, 47(2), 263 – 291.

[5] Hersh Shefrin, (2002) "Beyond Greed and Fear: Understanding behavioral finance and the psychology of investing". Oxford University Press.

[6] Andrea Berdondini, "The Professional Trader's Paradox", (November 20, 2018). Available at SSRN: https://ssrn.com/abstract=3287968.

[7] Andrea Berdondini, "Application of the Von Mises' Axiom of Randomness on the Forecasts Concerning the Dynamics of a Non-Stationary System Described by a Numerical Sequence" (January 21, 2019). Available at SSRN: https://ssrn.com/abstract=3319864 or http://dx.doi.org/10.2139/ssrn.3319864.

[8] Andrea Berdondini, "Description of a Methodology from Econophysics as a Verification Technique for a Financial Strategy", (May 1, 2017). Available at SSRN: https://ssrn.com/abstract=3184781.

# Statistics the science of awareness

Andrea Berdondini

ABSTRACT: The uncertainty of the statistical data is determined by the value of the probability of obtaining an equal or better result randomly. Since this probability depends on all the actions performed, two fundamental results can be deduced. Each of our random and therefore unnecessary actions always involves an increase in the uncertainty of the phenomenon to which the statistical data refers. Each of our non-random actions always involves a decrease in the uncertainty of the phenomenon to which the statistical data refers.

## Introduction

This article proves the following sentence:
"*The only thing that cannot be created randomly is knowledge*"

## A true story of a true coincidence

Ann is a researcher, is a clever and beautiful researcher, one day she decides to do the following experiment: she wants to understand if she has some special abilities that allow her to extract the number 1 from a bag containing one hundred different numbers mixed in a random way.

Day 1, Ann takes the bag and randomly pulls out a number. The drawn number is not 1, so she failed, the drawn number is put back into the bag.

Day 2, Ann takes the bag and randomly pulls out a number. The drawn number is not 1, so she failed, the drawn number is put back into the bag.

……………

……………

The days pass and Ann fails every attempt but continues his experiment.

……………

……………

Day 100, Ann takes the bag and randomly pulls out a number. The number drawn is 1, so she is successful. The probability of finding number 1 by performing a random extraction is 1/100; this value represents an acceptable error that makes the result significant to support the hypothesis that the extraction is non-random. But Ann knows that this probability does not represent the uncertainty of her success, because this value does not take into account previous attempts. Therefore, she calculates the probability of randomly extracting the number 1, at least once, in a hundred attempts. The probability value thus calculated is 63%, since this value is very high, she cannot consider the success obtained as significant statistical data to support the hypothesis that the extraction is non-random.

Ann concludes the experiment and deduces, from the results obtained, that she has no special ability and the extractions are all random.

John is a data scientist, one day he is entrusted with the following task: he has to develop an algorithm capable of predicting the result of an experiment whose result is determined by a value from 1 to 100.

Day 1, an incredible coincidence begins, at the same time that Ann pulls a number John tests his own algorithm. The number generated by the algorithm does not coincide with the result of the experiment, so he failed.

Day 2, the incredible coincidence continues, at the same time that Ann pulls a number John tests a new algorithm. The number generated by the algorithm does not coincide with the result of the experiment, so he failed.

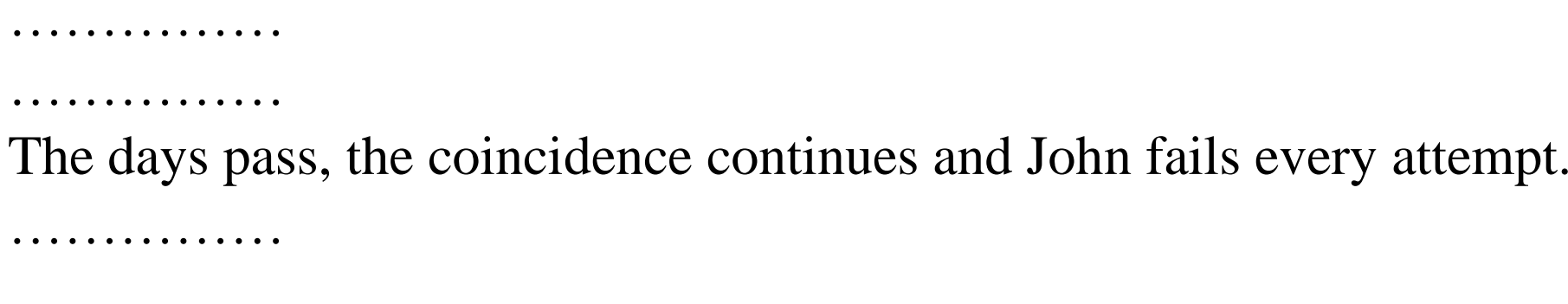

……………

……………

The days pass, the coincidence continues and John fails every attempt.

……………

……………

Day 100, the incredible coincidence continues, at the same time that Ann pulls a number John tests his new algorithm. The number generated by the algorithm coincides with the result of the experiment, so he is successful. The probability of predicting the result of the experiment by running a random algorithm is 1/100; this value represents an acceptable error that makes the result significant to support the hypothesis that the algorithm used is non-random.

For this reason, John writes an article in which presents the result obtained. The article is accepted, John is thirty years old and this is his hundredth article.

## Awareness breeds awareness

We call "researcher" a person who knows only his own attempts regarding the study of a certain phenomenon.

We call "reviewer" a person who does not actively participate in the study of a particular phenomenon but knows every single attempt made by each researcher.

Researcher 1: develops an algorithm that obtains the result R1 with respect to a phenomenon F. The probability of getting a result equal to or better than R1 in a random way is 1%.

Researcher 2: develops an algorithm that obtains the result R2 with respect to a phenomenon F. The probability of getting a result equal to or better than R2 in a random way is 1%.

Reviewer: defines a new result $RT = R1 \cap R2$. The probability of getting a result equal to or better than RT in a random way is 0.01%. Consequently, the uncertainty of the result RT is 0.01%.

## The absence of awareness reduces awareness

We call "researcher" a person who knows only his own attempts regarding the study of a certain phenomenon.

We call "reviewer" a person who does not actively participate in the study of a particular phenomenon but knows every single attempt made by each researcher.

Researcher 1: develops an algorithm that obtains the result R1 with respect to a phenomenon F. The probability of getting a result equal to or better than R1 in a random way is 1%.

Researcher 2: develops an algorithm that obtains the result R2 with respect to a phenomenon F. The probability of getting a result equal to or better than R2 in a random way is 100%.

Reviewer: defines a new result $RT = R1 \cap R2$. The probability of getting a result equal to or better than RT in a random way is 2%. Consequently, the uncertainty of the result RT is 2%.

# Quantitative trading

In this chapter, we will apply the statistical concepts discussed previously in the development of a quantitative trading system.

The first article "Description of a method of econophysics as a technique for verifying a financial strategy" deals with all the fundamental topics in the development of a quantitative trading system. Indeed, in addition to evaluating a trading strategy, this article also talks about many other aspects, such as the development of control methods on a running strategy.

The successive articles concern the non-ergodicity of the financial markets. In this situation, the data can be non-independent, so their statistical significance is difficult to define. Under these conditions, the risk associated with a trading algorithm is defined by its ability to predict the evolution of the system. Whenever a correct forecast is made on an evolution of the system, this forecast generates data that is independent of the previous data. From a statistical point of view, the independence of the data determines a lower probability of obtaining an equal or better result randomly. Consequently, this probability value is indicative of the number of correctly predicted system evolutions. As we have seen in the statistics articles, this parameter also represents the uncertainty of the statistical data.

In the last article, the Binomial evolution function is described, the purpose of which is to define a simple method to evaluate a trading algorithm. This function applies all the concepts described previously, allowing us to understand whether our trading algorithm is operating randomly or deterministically.

# Description of a methodology from Econophysics as a verification technique for a financial strategy

Andrea Berdondini

ABSTRACT. *In this article, I would like to draw attention to a method inspired by the analysis of stochastic models typical of quantum physics, and to utilise it to test a financial strategy. We start from the principal question asked by anyone involved in financial investments: Are the results obtained due to a correct interpretation of the market, or are they merely fortuitous? This is a plain question and it can be given an equally clear answer. The results obtained are due to a correct interpretation of the market if the probability of obtaining equal or better results randomly is very small (i.e. tends to zero as the number of times the strategy is used increases).*

## Description of the methodology

The logic underlying this method is very simple in essence: it consists in calculating the probability of obtaining the same results randomly. As we will see in the examples below, this technique is applied not only to results from a trial phase: it is also applied as a control method when the strategy is used on a real trading account.

In what follows, I present a short logical proof of the soundness of this method. The term 'soundness' was introduced by the famous mathematician David Hilbert and is used to indicate the absence of any contradiction within a mathematical logical proof. Indeed, contradiction is one of the main defects of methods of analysis based on equity line (performance) assessment.

The short demonstration I'm going to outline is based on two fundamental axioms:

1) Whenever we understand any kind of deterministic market process, the probability of our financial operation being successful increases by more than 50% (Von Mises' axiom of disorder from the early 1920s).

2) The probability of randomly obtaining a result that has been obtained through cognitive awareness of a deterministic market process tends to zero as the number of times the strategy is used increases.

The first axiom is derived from the famous "axiom of randomness" (or the 'principle of the impossibility of a gambling system') formulated by the mathematician Von Mises, whose original definition I quote: "the essential requirement for a sequence to be defined as random consists in the complete absence of any rules that may be successfully applied to improve predictions about the next number".

As a consequence of the two axioms given above, any correct market analysis will always tend to increase the probability of our prediction beyond the 50% mean, and this results in a consequent decrease in the probability of obtaining the same result randomly.

I will demonstrate this to you with a simple example. Suppose we are playing heads or tails with a rigged coin that gives us an above-50% probability of winning (let's say it's 60%). What is the probability of losing out after 10 coin tosses? Approximately 16.6% ...and after 50 tosses? Approximately 5.7% ...and after 100 tosses? Approximately 1.7%. As you can see, the probability tends to zero, and here the rigged coin represents a financial strategy that is implementing a correct market analysis.

By basing our method of assessment specifically on the calculation of this probability, we develop a method that is by definition free of contradictions. The absolute value of the probability turns out to be a sound estimate of the validity of our strategy.

The term 'deterministic process' which I used during the proof refers to the utilisation of a correct financial strategy, definable as the identification of a deterministic and non-random component that regulates the system we are studying (in our case, a financial market).

Methods based on studying the equity line may produce a positive outcome and at the same time have a 50% probability of obtaining the same amount of profit by chance. In this way, such methods lead to a contradiction, given that obtaining the same outcome randomly implies the absence of a cognitive process, which is just what is meant by assuming a "correct interpretation of the market".

These kind of methods are often based on the market stationary hypothesis (ergodic hypothesis). This hypothesis is considered by many experts not correct, on this topic have been written many articles the most famous is that written by the Nobel prize for physics Gell-Mann [1], two other interesting articles on this topic are [2], [3].

Figure 1 shows an equity line obtained with a purely random strategy. The algorithm is defined as follows: Each day you toss a coin to decide whether to open a buy position on the Nasdaq Index. The interval time chosen is from 1 January 2016 to 31 July 2017. If the position is opened, you toss another coin the next day to decide whether to close it or leave it open. As you can appreciate, this strategy functions in a completely inane and random way. Nevertheless, the equity line achieved is satisfactory: indeed, if we calculate the probability of obtaining an equivalent or better result randomly, we get a probability of approximately 50%. We therefore know that the result is void of significance, in spite of the equity line.

To conclude, it follows that the parameter to be linked to the validity of a financial strategy is not its performance but its statistical property of generating non-reproducible results in a random way.

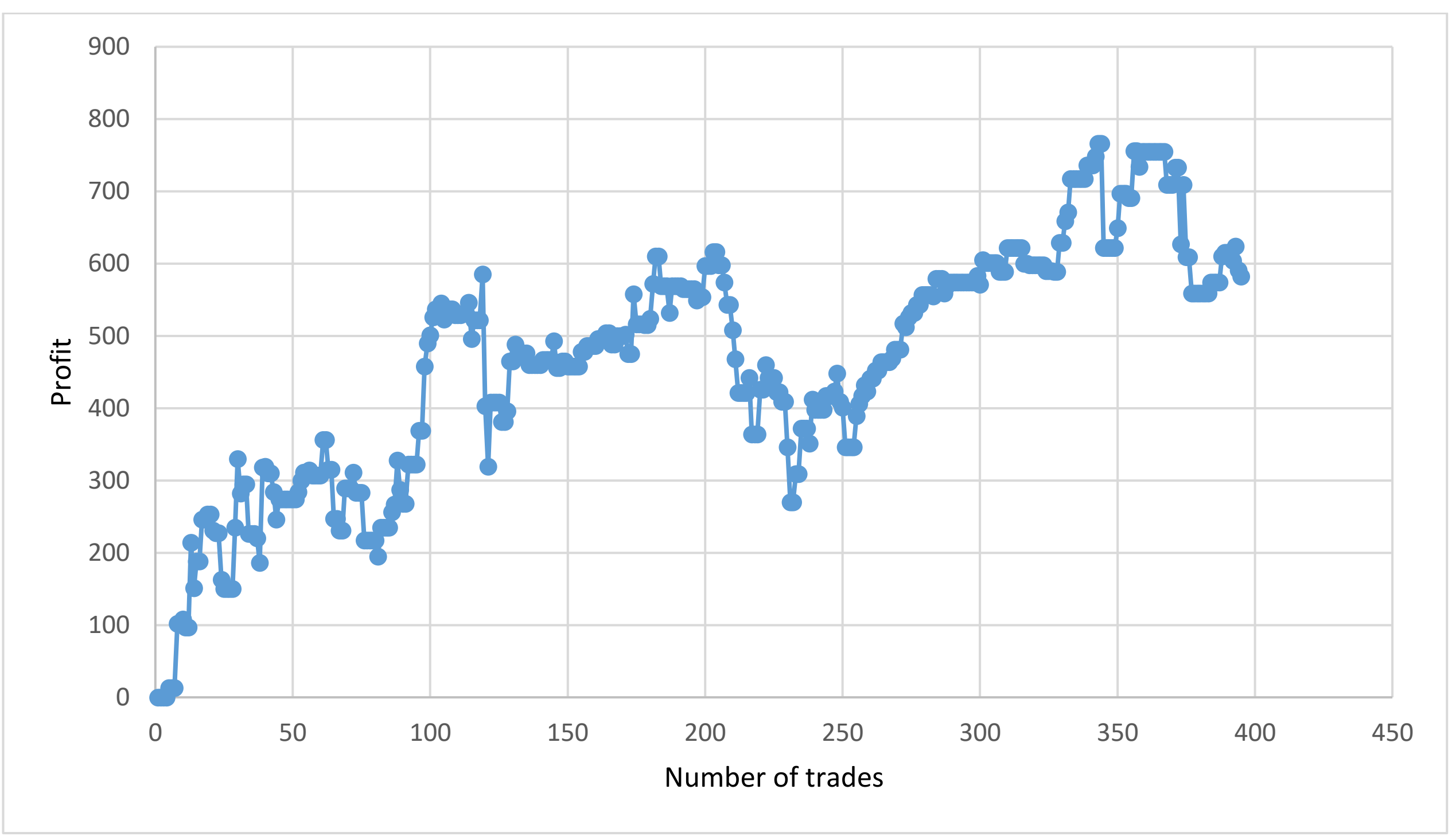

Figure 1: shows the equity line of a purely random trading system in which operations are opened and closed by simulating the toss of a coin.

## Description of the techniques used in operational practice of this verification method

How does one calculate the probability of randomly generating an equivalent or better performance? There are two ways: the first (and more precise) is to estimate this probability using the Monte Carlo method. The accuracy of this method is linked to the number of times we carry out the simulation. Its strength is the ability to obtain very precise values: its drawback is due mainly to the long calculation times required to obtain the estimate of probability.

The second method (which I have identified) uses exact formulae taken from statistics; each formula is applied to a particular class of random variables. Unfortunately, financial operations do not fall under any type within this class of variables. This problem is solved by applying a transform to the financial operations, which renders them suitable for the chosen analytical formula.

This transform adds an error to the calculation of our probability, but it has the advantage of being calculable in one single equation and therefore without requiring massive computational resources, as do the Monte Carlo methods.

## Use of the method as a control parameter of a strategy

This method is utilised not just during the test phase, but is also extremely useful as a way of monitoring the trading system. Each time we carry out an operation, we update the probability value for obtaining that result randomly. We do not calculate this probability across the whole sample of operations conducted, but extrapolate it from to the *N* most recent operations. The optimal value for *N* depends on the strategy, and in particular on the frequency with which operations are conducted over time. Once *N* has been set, we calculate our probability and compare it with a probability we have established that represents the level of risk we take to be acceptable (this definition of risk will be explained in a separate section). If the probability value exceeds the parameter set by us, the trading system locks itself and continues trading in virtual mode only. When the probability falls below the threshold parameter we have set, the trading system resumes actual trading. In this way, trades are effected only when the market is understood, and the trading system is blocked when we are operating in a regime considered to be random.

This method is much more efficient than the usual performance-based methodologies; such methods carry the risk of allowing themselves to incur unnecessary losses. It may happen with this method that the strategy is blocked even when trading at a profit, given that a random regime has a 50% probability of success.

Having said this, obviously a trading system will have its internal performance controls, but their purpose is purely to monitor for possible system crashes or any programming bugs.

What we have described thus far can be fine-tuned. A characteristic of all good quantitative trading systems is to be capable of operating even at high frequencies while leaving unchanged the logical schema on which the trading system is based. This enables us to run a trading system solely as a method of monitoring (hence in virtual mode), at a very high frequency of operations, and to obtain thereby a much more numerous statistical sample in less time, increasing the reactivity of our method of control.

## Use of the method to in the process of developing a financial strategy

This approach has another great merit, which is to help us direct our research in the right direction. Let us assume we develop two strategies:

1) The first has profits on a historic series of 10% annuities, but with a very high probability of obtaining the same results randomly;

2) a second, instead, has low profits of 1%, but with an extremely low probability of obtaining the same results randomly.

It is perfectly obvious that, if we follow the theory we have expounded, we will discard the first strategy, as there is a very high probability that the 10% profit has been obtained by mere chance. The second strategy yields low profits but the low probability value obtained means we are on the right track for understanding a deterministic and non-random market process (which, if studied more closely, could lead to more profitable financial strategies).

If we had not applied this method, we might have thought that the first strategy (with higher gains) was the right one. But this would have been at risk of losing money over the medium to long term. We would have ended by discarding the second strategy and missing an opportunity to study and understand something important we had sensed about the market.

## Example of the use of this methodology

I report the following practical example; the figure shows the trend of a hypothetical stock. The value of this stock on a thousand time intervals rises one unit 60% of the time and goes down by one unit 40% of the time. In order to simplify the calculations the price movement is unitary. Now let us look at two strategies that execute 500 trades, each trade lasts an interval of time. The first strategy execute only buys and in order to choose when to buy, flips a coin, if it win opens a trades if loses it waits for the next unit of time and repeats the operation. The second strategy, on the other hand, is a strategy that sells only, but does not do it in a random way, it uses information that allows it a 10% advantage in determining the drops of the value stock.

The first strategy gets a profit of 100 by winning 60% of trades. Now we calculate the probability of obtaining a better result in a random way, to do this we use the binomial cumulative distribution function with the following parameters:

p = 60% (probability of win)

k > 300 (number of wins)

L = 500 (total number of tests)

<u>The probability of victory is 60% because in the graph shown the value of the stock rises by one unit 60% of the time and goes down by one unit 40% of the time.</u>

Using these data, we get a probability to get better results randomly of the 48.3%.

Now let us consider the second strategy, this strategy has a total result of zero in practice it performs 250 winning trades and 250 losing trades (the 10% advantage allows it to increase the probability of victory from 40% to 50%). Also in this case, we calculate the probability of obtaining better results randomly, to do this we use the formula of the binomial cumulative distribution function with the following parameters:

p = 40%

k > 250

L = 500

Using these data, we get a probability to get better results randomly of the $2.5 \cdot 10^{-4}$ %.

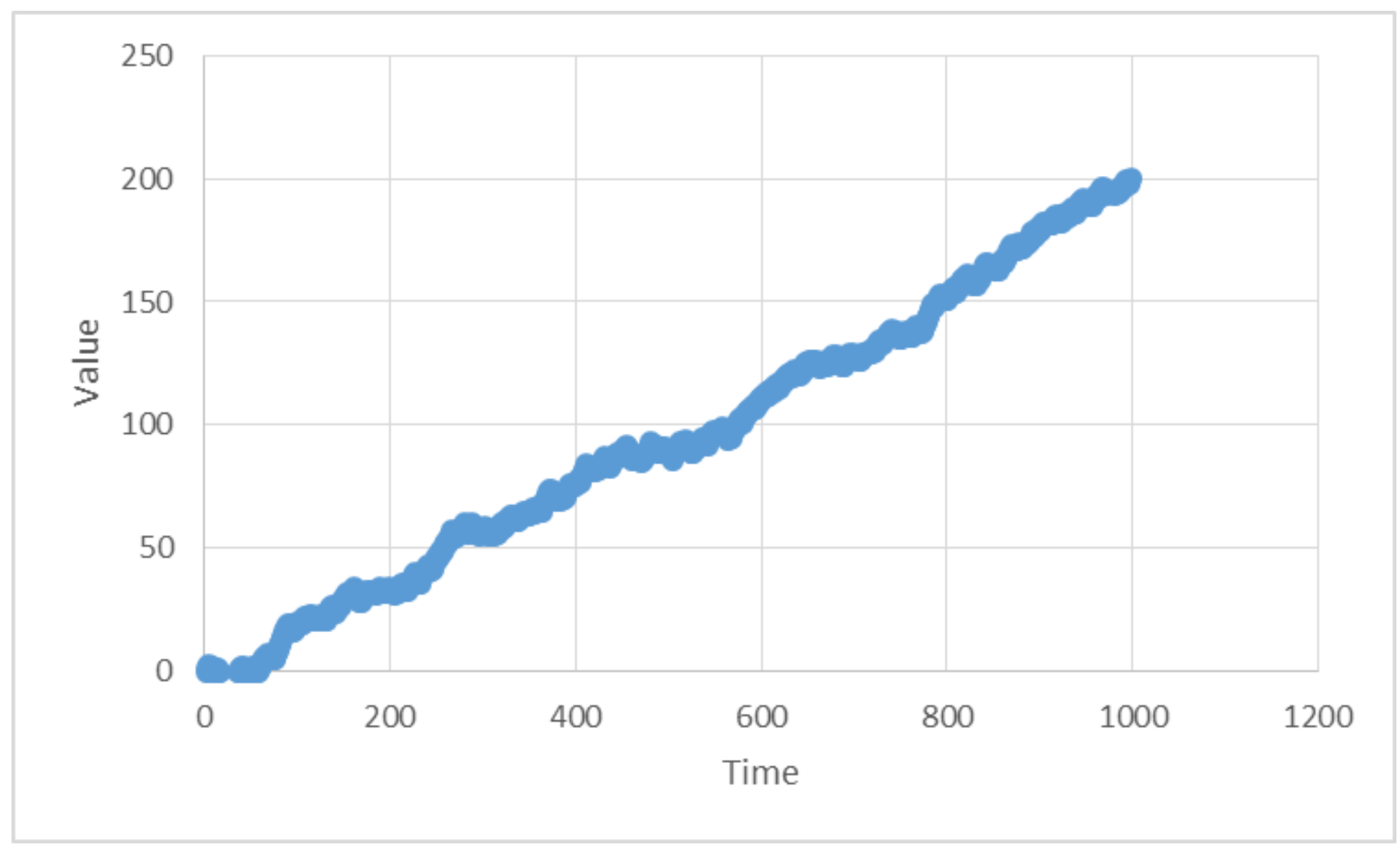

Figura 2: trend of a stock that rises 60% of the time and falls the remaining 40%.

Analyzing these results it is clear that the profit of 100 made by the first strategy is not significant and this evaluation is correct because we know that the strategy flip a coin in order to decide whether to open a buy position. The second strategy does not make any profit but such a low probability makes this result significant and even this evaluation is correct, in fact we know that this strategy implements useful information that allows it to increase the probability of victory by 10%.

What would have happened if we had used the competition evaluation method? We will have discarded the second strategy and reward the first strategy, a completely random strategy. Where is the error? The error derives from considering result as an always useful element for future profits instead the fundamental element to be winning is to be able to understand the rules to which a system is subject, in this case the financial market. This knowledge allows us to act in a non-random way and this feature can be detected only if we calculate the probability of obtaining better results at random.

## A new definition of risk

There are many definitions of risk in the financial field: risk is in any case seen as the probability (uncertainty) of incurring a given loss. If we recall the statistical example given above of tossing the rigged coin, we can see how risk is intimately linked to our understanding of the market and as such how it tends to zero the more times we repeat the statistical experiment described above.

The value of this probability can never be zero: think, for example, of the actions we perform in our daily lives, actions that all have a certain level of risk – understood as the probability of bringing about a negative event. We take these actions into consideration nevertheless, because we know that the risk associated with them is so low as to be statistically acceptable – for example the risks associated with travelling by plane.

It therefore becomes extremely important to implement methods that evaluate the validity of our strategy in a sound way, so that we can estimate risk and plan the investment correctly.

## Conclusion

In this article, I wanted to draw your attention to a different way of viewing the performance of a trading system, a way that is not bound to its absolute value, but linked to one of its statistical properties. As I demonstrated in the first section, this involves well defined behaviours when we operate with cognitive awareness on the market. The approach is a fundamental one because it recognises the high likelihood of being successful on financial markets, even over long periods, in a completely random way. Let's not forget that financial markets have only two possible directions. This implies that, even fortuitously, there is a 50% chance of making the right choice. Furthermore, such trends can continue for years. Therefore, it is crucial to look away from the profit line and to appreciate in a rigorous and scientific way whether our strategies are the product of chance or of a true understanding of the market. One thus trades only if one understands the market, thereby actually reducing the element of fortuitousness. Investing in this way has nothing to do with chance but becomes cognitively aware and more secure.

Gambling is defined as follows:

"*The gaming activity in which profit is sought after and in which winning or losing occurs predominately by chance, skill being of negligible importance*"

From this definition, it follows that if the element of fortuitousness is not factored into the investment decision-making process, it is never possible to prove that money invested is free of exposure to chance, and therefore to uncontrolled risk. The calculation of probability illustrated above therefore becomes an essential and irreplaceable requirement for bringing investment out of the area of gambling, making it more cognitively aware, and therefore less risky.

## References

[1] O. Peters, M. Gell-Mann. “Evaluating gambles using dynamics”, Chaos: An Interdisciplinary Journal of Nonlinear Science, 2016; 26 (2): 023103 DOI: 10.1063/1.4940236

[2] O. Peters, “Optimal leverage from non-ergodicity,” Quant. Finance 11,1593–1602 (2011).

[3] O. Peters and A. Adamou, “Stochastic market efficiency,” preprint arXiv:1101.4548 (2011).

# The professional trader's paradox

Andrea Berdondini

ABSTRACT: In this article, I will present a paradox whose purpose is to draw your attention to an important topic in finance, concerning the non-independence of the financial returns (non-ergodic hypothesis). In this paradox, we have two people sitting at a table separated by a black sheet so that they cannot see each other and are playing the following game: the person we call A flip a coin and the person we'll call B tries to guess the outcome of the coin flip. At the end of the game, both people are asked to estimate the compound probability of the result obtained. The two people give two different answers, one estimates the events as independent and the other one considers the events as dependent, therefore they calculate the conditional probability differently. This paradox show how the erroneous estimation of conditional probability implies a strong distortion of the forecasting skill, that can lead us to bear excessive risks.

## The professional trader's paradox

In order to explain how much danger is considering the financial returns as independent, I want to present to you this paradox. We have two people sitting at a table separated by a black sheet so that they cannot see each other and are playing the following game: the person we call A flip a coin and the person we'll call B tries to guess the outcome of the coin toss. This game lasts an arbitrary time interval and the person A has the freedom to choose how many tosses to make during the chosen time interval, the person B does not see the coin toss but can at any time, within the time interval, make a bet. When he makes a bet if he guesses the state the coin is in now, he wins. The person A decides to make a single coin flip (just at the beginning of the game) we say that the result is head, the person B decides within the same time interval to make two equal bets, betting both times on the exit of the head. The result is that B made two winning bets.

Now we ask ourselves this question: what is the correct compound probability associated with the result of this game? Let us ask this question to the person B who answers: every time I had bet I could choose between head and cross so I had a 50% chance of winning the bet; I won two bets so the compound probability is $0.5 \cdot 0.5 = 25\%$. Now let us say the same question to A the person who flip the coin, he replies: the probability is 50% I have flip the coin only one time within the defined time interval, so its prediction probability cannot be higher at 50%. The fact that the other player has made two bets has in practice only divided a bet in two is a bit 'as if to the racecourse we are made two distinct bets on the same horse on the same race, this way of acting does not increase the forecasting skill. Both answers seem more than reasonable, but as every mathematical paradox, the two answers contradict each other. At this point, will you ask yourself which of the two answers is correct?

We can resolve this paradox using the mathematical formula of the compound probability:

$P(E1 \cap E2) = P(E1 \mid E2)\, P(E2) = P(E2 \mid E1)\, P(E1)$.

The probability that both events (E1, E2) occur is equal to the probability that E1 occurs P(E1) multiplied by the conditional probability of E2 given E1, $P(E2 \mid E1)$.

Seeing the formula, we immediately understand that the difference in response given by A and B is due to the different estimation of conditional probability $P(E2 \mid E1)$. Person B estimates the conditional probability in this way P(E2 | E1) = P(E2) treating the events as completely independent, while person A estimates the conditional probability in this other way P(E2 | E1) = 1 treating the events as completely dependent.

Which of the two answers is correct? The right answer is given by the person who has the knowledge to correctly estimate the conditional probability P(E2 | E1) and between the two players only the person that flip the coin can correctly estimate the conditional probability. Player B, on the other hand, not being able to see A that flip the coin, therefore he does not have the necessary information to estimate this probability correctly. Another way to understand this result can be found analysing the following question: what is the probability in this game of winning twice in a row by betting both times on the head?

The answer to this question is not always the same but it depends if after the first bet the person that flip the coin performs a new launch or not. If you make a new launch, the probability is $0.5 \cdot 0.5 = 25\%$ if instead as in the case of this paradox no further coin flip is performed the probability is $50\%$. So, in order to answer correctly, you need to know the number of launch made and this information is knows only from the person (A) that perform the coin flip and he's the only one can be correctly calculate the conditional probability.

If we bring this paradox on the financial markets, we understand that player A represent the financial instruments and player B represent the traders who try to beat the market. This gives us an extremely important result: all the traders make the same mistake, doing the same thing that player B did in this paradox. They consider their trades as completely independent of each other and this involves as we have seen, a strong distortion of the forecasting skill that can lead the traders to acquiring a false security that may lead them to bear excessive risks.

Player B, like the traders, think that the statistical information about his forecasting skill depends on his choice (I choose head instead of cross, I buy instead of selling) this is a big mistake because this statement is true only when these kinds of bets are independent of each other. In practice, this statement is true only when I place a bet by event in this case, the results are independent of each other and therefore these bets have a statistical meaning.

The problem is that in everyday life this equivalence is always respected. Therefore, our brain considers this equivalence always true so when we make trading we mistakenly consider our operations as independent despite the statistical evidence of non-independence (non-normal distribution of the results).

## Conclusion

In this short article, I wanted to introduce one of the most important topics in finance, which concerns the non-independence of the results. Considering the financial returns as independent is equivalent to considering the financial markets stationary (ergodic hypothesis).

This hypothesis is considered by many experts not correct, on this topic have been written many articles [1], [2], [3]. What is the reason why such significant statistical evidence has been ignored, the main reason is the total lack of methods able to estimate the conditional probability P (A | B).

In my previous article [4] I have explained an innovative method used in order to evaluate a financial strategy under the condition of the market non-stationary hypothesis (non-ergodic hypothesis). This approach is based on the axiom of disorder (von Mises), this mathematical axiom applied on financial markets can be enunciated in this way:

"*Whenever we understand any kind of deterministic market process, the probability of our financial operation being successful increases by more than 50%*" (Von Mises' axiom of disorder from the early 1920s).

As a consequence of this axiom given above, any correct market analysis will always tend to increase the probability of our prediction beyond the 50% mean, and this results in a consequent decrease in the probability of obtaining the same result randomly. To conclude, it follows that the parameter to be linked to the validity of a financial strategy, is not its performance but its statistical property of generating non-reproducible results in a random way.

I will demonstrate this to you with a simple example. Suppose we are playing heads or tails with a rigged coin that gives us an above-50% probability of winning (let's say it's 60%). What is the probability of losing out after 10 coin tosses? Approximately 16.6% ...and after 50 tosses? Approximately 5.7% ...and after 100 tosses? Approximately 1.7%. As you can see, the probability tends to zero, and here the rigged coin represents a financial strategy that is implementing a correct market analysis.

Now we return to the paradox that I exposed and we note how the presence of a dependence between the first and the second bet has modified the conditional probability of the second one from 0.5 to 1. This increase of the conditional probability has the consequence that the result of the second bet can be obtained randomly.

In fact, if we move the second bet randomly within the time interval from the first bet to the second one, the result is always the same because the player who flip the coin (player A) does not execute other coin tosses in this time interval. Consequently, the second bet cannot be considered to evaluate the forecast skill. Therefore, considering a system not stationary involves a reduction of the number of events to be considered for a statistical evaluation so if a data set proves to be statistically significant under the condition of stationarity of the system, the same data set may no longer be statistically significant if the system is considered non-stationary.

## References

[1] J. Barkley Rosser Jr. 'Reconsidering ergodicity and fundamental uncertainty'. In: Journal of Post Keynesian Economics 38 (3 2015), 331–354. doi: 10.1080/01603477.2015.1070271 (1).

[2] M. Sewell. "History of the efficient market hypothesis". Technical report, University College London, Department of Computer Science, January 2011.

[3] O. Peters, M. Gell-Mann. "Evaluating gambles using dynamics", Chaos: An Interdisciplinary Journal of Nonlinear Science, 2016; 26 (2): 023103 DOI: 10.1063/1.4940236.

[4] Berdondini, Andrea, "Description of a Methodology from Econophysics as a Verification Technique for a Financial Strategy ",(May 1, 2017). Available at SSRN: https://ssrn.com/abstract=3184781.

# Application of the Von Mises' axiom of randomness on the forecasts concerning the dynamics of a non-stationary system described by a numerical sequence

Andrea Berdondini

ABSTRACT: In this article, we will describe the dynamics of a non-stationary system using a numerical sequence, in which the value of terms varies within the number of degrees of freedom of the system. This numerical sequence allows us to use the Von Mises' axiom of randomness as an analysis method concerning the results obtained from the forecasts on the evolutions of a non-stationary system. The meaning of this axiom is as follows: when we understand a pattern about a numerical sequence, we obtain results, intended as forecast on the next sequence number, which cannot be reproduced randomly. In practice, this axiom defines a statistical method capable of understanding, if the results have been obtained by a random algorithm or by a cognitive algorithm that implements a pattern present in the system. This approach is particularly useful for analysing non-stationary systems, whose characteristic is to generate non-independent results and therefore not statistically significant. The most important example of a non-stationary system are financial markets, and for this reason, the primary application of this method is the analysis of trading strategies.

## Introduction

The first problem that must be faced, in order to apply the axiom of randomness of Von Mises as an analysis method, is to be able to describe the evolution of a system using a numerical sequence. For this purpose, we use a sequence in which the value of terms can change within the number of degrees of freedom of the system. Then we also introduce a temporal progression represented by a series of increasing integers. With these two sequences, we can characterize a dynamic that describes the evolution of the system that we are studying. In this way, we can apply the Von Mises' axiom of randomness, which defines the statistical characteristic of a random sequence. The axiom is as follows: "*the essential requirement for a sequence to be defined as random consists in the complete absence of any rules that may be successfully applied to improve predictions about the next number*".

This axiom tells us that when we understand a pattern about a numerical sequence, we can obtain results, intended as forecast on the next sequence number, which cannot be reproduced randomly. Knowing that the values of our numerical sequence represent the dynamics of a system, we have found a method to evaluate the results. Therefore, if the probability of obtaining equal or better results randomly is very small (i.e. tends to zero as the number of times the strategy is used increases), it means that the forecasting method uses rules present in the system. Consequently, we can use the developed method in order to predict the evolution of the system; this approach is fundamental in the study of non-stationary systems. In fact, if we are in a condition of non-stationarity, the results can be non-independent, the consequence of which is a reduction in their statistical value. In order to explain such a complex topic in a simple way, I have created a paradox [1], in which two players challenge each other in a game, where is possible to move from a stationary to a non-stationary condition by changing the rules of the game. The consequence is that the data at the beginning are independent, and then once the changes are made, they become dependent on each other and therefore useless for statistical purposes.

For that reason, a large number of data obtained under these conditions, can be not statistically useful in order to understand if the method that produced the data is not random. In this case, the performance of the forecasting algorithm is no longer a reliable data; therefore we need to find a new statistical indicator in order to evaluate the results. In fact it happens very often in finance, that a strategy that despite having obtained excellent results by performing a numerous number of operations, suddenly stop working generating large losses, proving in this way to be a completely random strategy. There are many researches on the non-stationarity of the financial markets; some articles about this topic are [2], [3], [4], [5].

## Description of the dynamics of a non-stationary system by means of a numerical sequence

We start by considering a non-stationary system characterized by two degrees of freedom. We made this choice for two reasons: first in order to simplify the treatment, second among the systems with two degree of freedom there are the financial markets, that represent the primary application of the method that we will go to expose.

Then we associate to this system a succession of values that can vary between 1 and 2. In practice when on this system there is a change in the direction in which a deterministic process is acting, we have the result that the value of the succession change respect to the previous term. When we speak of "deterministic process", we indicate a deterministic and non-random force that acts on the system. We also introduce the simplification that the deterministic process acts as a constant force, without variations in intensity in the direction of the two degrees of freedom; this allows us to characterize the dynamics of the system with a single numeric succession.

Finally, we add a temporal metric to the system; we can do this by using a series of increasing integer numbers, in this way every value of my succession unambiguously corresponds to a value in my temporal progression.

The result obtained is the following: we have two sequences, one that describes the direction of the force acting on the system F and the other one that describe the progression of the time T:

**F** 1 1 1 1 1 1 1 2 2 2 2 2 1 1 1 1 1 1 1 2

**T** 1 2 3 4 5 6 7 8 9 10 11 12 13 14 15 16 17 18 19 20

Analysing these two sequences we can note that the force (deterministic process) that acts on the system changes direction three time, the first at time T = 8, the second at time T = 13 and the third at time T = 20. Figure 1 shows the dynamics described by the two sequences; the temporal progression has also been included in the graph.

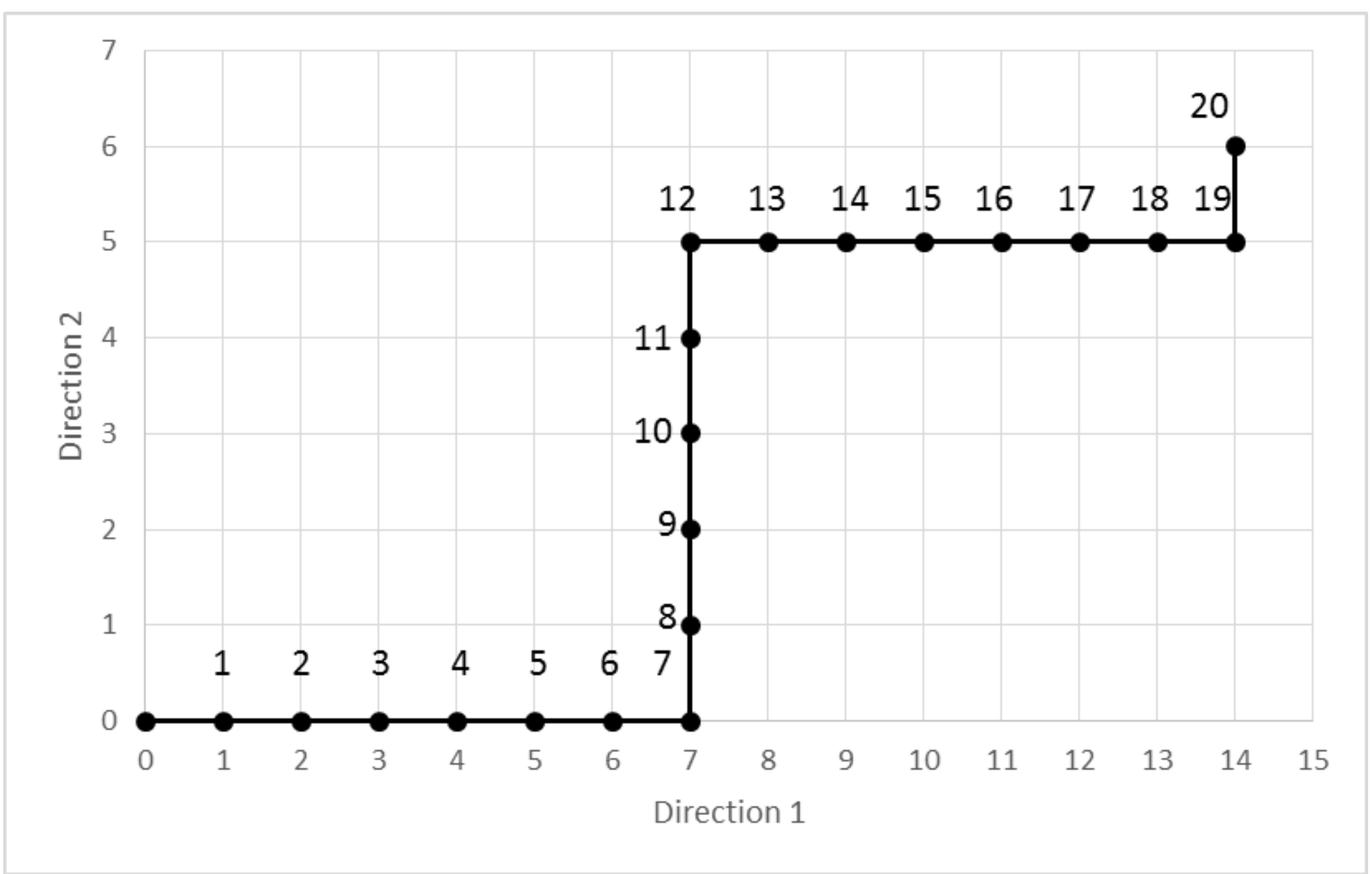

Figure 1: The dynamics of the system described by the two sequences.

## Application of the Von Mises' axiom of randomness as an analysis method concerning the results obtained by a forecasting algorithm

At this point, we are going to apply the Von Mises' axiom of randomness in order to evaluate the results obtained by a forecasting algorithm. According to the axiom, reported and analysed in the introduction of this article, if we have developed a non-random method that implements a pattern present in the system, we will be able to predict when the term of the succession varies respect the previous term.

This statement is crucial, because we are saying that only the forecasts concerning system changes are useful in order to understand whether the method used is random or not. In fact, if we make two forecasts about two terms of the succession at time 8 and 12, betting that the value of the succession is two, we obtain two successes. Are these two results independent of each other? This question is fundamental, because only if they are independents are useful for statistical purposes in order to evaluate the forecasting method.

In order to answer to this question we must use the formula of the compound probability that we report:

$$P\,(E1 \cap E2) = P\,(E1 \mid E2)\; P\,(E2) = P\,(E2 \mid E1)\; P\,(E1) \qquad (1)$$

The probability that both events (E1, E2) occurs, is equal to the probability that E1 occurs P(E1), multiplied by the conditional probability of E2 given E1 occurs P(E2 | E1).

Seeing formula (1) we understand that the correct calculation of conditional probability P (E2 | E1) depends on whether the events are independent or not. In fact, in the case of independent events P(E2 | E1) = P(E2), instead in the case of events completely dependent on each other P (E2 | E1) = 1 or 0.

In our example the probability P(E1) = 1/8. In fact, the probability of winning in a random way betting that the value of the sequence within the first 8 time positions is equal to 2 is 1/8, since in the first seven positions the sequence has value 1 and only at the eighth position has value 2. In practice, we can see this value like the probability of a random draw of a marble with the value 2 from a bag that contains eight marbles, seven of them with the value 1 and one with the value 2. Therefore, P (E1) represents the probability of obtaining the same result using a random strategy. In this way, we are able to derive a probability from a succession of data concerning the dynamics of the system.

Now we try to calculate the value of P (E2); studying the time interval from the ninth position to the twelfth position, we see that the system remains constant. Therefore, the probability of betting randomly on the value 2 and win, within of this time interval, it is equal to 1. This is because all positions have value two, so a random strategy obtains the same result with a probability of 100%. In this case, we can treat the second bet as completely dependent from the first bet. This means that we can consider the two bets, from the statistical point of view, like a single bet.

Now let's try to shift the second bet from the twelfth position to the twentieth position, in this case we will have P(E2) = 5/7. In fact, in the time interval from the ninth position to the twentieth position, there are seven values equal to 1 and five values equal to 2. With this variation in the second forecast, the compound probability will be equal to:

$$P(E1 \cap E2) = P(E1)\, P(E2 \mid E1) = P(E1)\, P(E2) = 1/8 \cdot 5/7 = 0.089$$

The two events in this case are independent of each other, so they are useful for statistical purposes in order to understand if the method used can predict the evolution of the system.

From these two examples, we can deduce the following conclusion: when we want to study a deterministic process, every time that it determines a change in the system, we have the possibility to execute a prediction on the evolution of the system, whose probability of success with a random strategy turns out to be minor 1. This involves a decrease in the compound probability whose meaning, according to the Von Mises' axiom of randomness, is to indicate the presence of a deterministic process that acts on the system.

In practice, whenever that a deterministic process changes the system status, we can detect it by making a prediction. Consequently, in order to detect a deterministic process with a low error, it is necessary that it has produced a statistically significant number of variations in the system. On which our forecasting method has carried out a large number of independent predictions. Therefore, every forecast must concern a single variation of the system.

This approach is crucial for non-stationary systems where the forecasts may be not independent. For that reason, it is fundamental to calculate the compound probability, which is the probability of obtaining the same results in a random way. We also remember that within the compound probability (1), there is the calculation of the conditional probability, in which it is taken into account if the events are dependent or independent from each other. In fact, the goal of this method is to discard all non-independent forecasts, whose contribution is to make me overestimate the forecast skill of the algorithm. The consequence of this erroneous evaluation can lead us, for example in the financial field, to bear excessive risks.

# Conclusion

The analysis of results, concerning the evolution of a non-stationary system, represents one of the most important problems of applied mathematics still unresolved. In this article, I propose the use of the Von Mises' axiom of randomness as a method of evaluating the data obtained under these conditions. This axiom, as explained previously, defines a statistical characteristic that assumes a well-defined behaviour when we operate consciously. In practice, the results generated by a non-random forecasting algorithm, that implements knowledge about the considered system, cannot be reproduced randomly. Consequently, the probability of obtaining equal or better results randomly tends to zero as the number of times the strategy is used increases. In this way, we shift our attention from the value of the result, which under the non-stationary condition may have been produced with non-independent forecasts, to its statistical characteristic correlated to its non-random behaviour.

In order to apply this method, we have defined a simple mathematical model whose task is to describe the dynamics of a system by means of a numerical sequence. Using this mathematical model, we have shown how to compute compound probability from a series of results obtained with a forecasting algorithm. Then, analysing some examples, we have deduced some important considerations. In particular, we have seen that when there is a change in the system, we have the possibility to make a prediction whose probability of success is less than 1. The consequence of this is a decrease of the compound probability whose meaning, according to the Von Mises' axiom of randomness, is to indicate the presence of a pattern on the system. Therefore, if we want to detect it with a low error we must have made a statistically significant number of independent predictions, concerning each a single variation of the system. In this way, the forecast algorithm proves to be able to predict future system evolutions. This characteristic, therefore, becomes fundamental in estimating the risk of a strategy that operates on a non-stationary system, such as the financial markets. In fact, a correct risk assessment must always take in considerations the forecast skill of the algorithm about the future evolutions of the system.

In the next article, we will introduce a purely random component within the dynamics of a non-stationary system. In this way the deterministic process will be replaced by a stochastic process described by a probability density function. This step is essential in order to apply this method in the financial field**.**

# References

[1] Berdondini Andrea, "The Professional Trader's Paradox", (November 20, 2018). Available at SSRN: https://ssrn.com/abstract=3287968.

[2] J. Barkley Rosser Jr. 'Reconsidering ergodicity and fundamental uncertainty'. In: Journal of Post Keynesian Economics 38 (3 2015), 331–354. doi: 10.1080/01603477.2015.1070271 (1).

[3] M. Sewell. "History of the efficient market hypothesis". Technical report, University College London, Department of Computer Science, January 2011.

[4] O. Peters, M. Gell-Mann. "Evaluating gambles using dynamics", Chaos: An Interdisciplinary Journal of Nonlinear Science, 2016; 26 (2): 023103 DOI: 10.1063/1.4940236.

[5] Berdondini, Andrea, "Description of a Methodology from Econophysics as a Verification Technique for a Financial Strategy ",(May 1, 2017). Available at SSRN: https://ssrn.com/abstract=3184781.

# Binomial evolution function

## (The only function you need to evaluate a trading strategy)

Andrea Berdondini

ABSTRACT: A trading strategy represents a forecasting model and a forecasting model will generate profits in the future if it has demonstrated in the past that it can predict with a good probability of success the evolutions of the system on which it operates. Since the theory is so simple, why is it so difficult to evaluate a trading strategy? The reason is the following: in finance, we think we see the evolutions of the system but we are only observing the random component due to unpredictable events. Instead, the deterministic component that represents the true evolution of the system is not visible. Unfortunately, not seeing the deterministic component leads to a very important statistical problem: if I execute 10 winning predictions, I do not know if these operations concern 10 evolutions of the system or concern only one evolution of the system. Therefore, I do not know, if my forecasting algorithm acted intelligently by predicting 10 different evolutions or if it acted irrationally by dividing a bet into 10 smaller ones on the same evolution. Distinguishing between these two scenarios is essential to correctly evaluate the trading strategy and this discrimination can be done by understanding whether the trades are independent of each other. In fact, the condition of independence is the only statistical way to be sure that the trades concern forecasts of different evolutions. The Binomial evolution function solves this problem by transforming the succession of trades into a succession of independent results.

## Introduction

The key to understanding whether an investment strategy can generate profits in the future is to shift the focus from the simple result (profit/loss) to the ability of the algorithm to predict the evolution of the market, under the assumption that markets are non-stationary [1][2]. Being "non-stationary" implies that operations are dependent on each other (market dynamics change over time), and consequently the data collected could lose statistical significance if treated as independent. When dependencies unknown to us are created, traditional statistical analysis risks leading to incorrect conclusions.

**Consequently, the main goal of any function that wants to make a qualitative analysis about a trading algorithm is to transform the results into a series of independent data.**

The Binomial evolution function solves this problem by transforming the succession of trades into a succession of independent outcomes. The data are then analyzed using the binomial distribution that allows us to calculate the probability of obtaining an equal or greater number of successes through a purely random process. The probability value obtained in this way turns out to be a statistically correct evaluation of the investment strategy.

## Transformation of the Trade Series

The first step in obtaining a sequence of independent trades consists in applying the following transformation: if we have N forecasts on a system with K degrees of freedom, our predictions must be transformed into a series in which each **forecast "bets" on a different evolution** from the one immediately preceding it.

In the case of financial markets, there are two degrees of freedom: **up (buy) or down (sell)**. Consequently, in our new series, there should never be two consecutive operations of the same type (two buys or two sells). Therefore, the sequence must alternate buy and sell. From a practical point of view, this transformation is applied by **inserting an opposite operation** between two consecutive trades in the same direction. If, for example, we have two consecutive buys with a time interval ΔT between the closing of the first and the opening of the second, a sell trade of duration equal to ΔT is introduced in between.

**Why this transformation?**

- If after a bullish phase (buy) we think that the trend has ended, inserting a test opposite operation (sell), for the period ΔT, allows us to test our hypothesis of “end of trend”.
- If the market actually remains bullish, this intermediate operation will have a probability of loss greater than 50%. If, on the other hand, our assumption is correct and the bullish trend is over, the intermediate sell trade will have a probability of success that approaches 50%.

In conclusion, **each intermediate operation inserted tests the hypothesis that the two consecutive operations (in the same direction) are truly independent.**

## Convert to a binary sequence

After creating the new sequence of operations (all alternating), **we proceed to convert it into a binary sequence**:

- We assign the value 1 to the trades closed in profit.
- We assign the value 0 to the trades closed in loss.

This additional step is necessary to **not overestimate the impact of some individual results** that may depend on random market events. By making each operation equal, we reduce the effect of large gains or isolated losses.

## Selection of Independent Trades

In this phase, we identify the trades that we are sure are independent. In practice, this is done by selecting only the trades that were executed on different market evolutions. In this way, each trade represents an independent forecast with respect to the previous one.

To make this selection, we use the binary sequence of 0 and 1 obtained in the previous steps, where each symbol represents the result of a trade (for example, 1 for profit and 0 for loss). If a 0 is preceded by another 0, or a 1 is preceded by another 1, then we can be sure that the second trade refers to a different evolution of the market. This is because the system, by construction, generates alternating forecasts (buy/sell). Therefore, obtaining two consecutive equal results implies that the market must have changed.

**Consequently, the principle on which this method is based is the following: two equal and opposite predictions cannot give the same outcome if the system has remained unchanged.**

Conversely, if a 0 is preceded by a 1 or a 1 is preceded by a 0, the two results may be compatible with a stationary market situation. In this case, we cannot be sure that the two trades refer to distinct evolutions, so we consider them potentially dependent and exclude them from further analysis.

**At the end of this process, only independent trades remain, those that we are sure will be executed on different evolutions of the system.**

Therefore, when we want to study a deterministic process, every time it determines a change in the system, we have the possibility of making a prediction on the evolution of the system whose probability of success randomly turns out to be less than 1. This leads to a decrease in the composite probability whose meaning, according to the axiom of disorder of Von Mises, is to indicate the presence of a deterministic process that acts on the system.

In practice, every time a deterministic process causes a change in the state of the system, it can be detected by making a prediction [3]. Consequently, a deterministic process to be detected with a low error must have produced a statistically significant number of variations in the system. Furthermore, our forecasting method must have made a sufficiently large number of independent predictions, each corresponding to a single evolution of the system.

### Randomness test using the Binomial distribution

Finally, we calculate the probability of obtaining an equal or greater number of successes through a **purely random process**, as if we were tossing a fair coin (probability 50%). For this purpose, we use the binomial distribution, where:

- k is the number of successes (trades with value 1).
- n is the total number of trades.
- p is equal to 50% (assuming fair coin).

If the probability of observing similar or better results randomly is very low, it means that the algorithm has a good chance of being deterministic (able to predict the market). On the contrary, if the results can be obtained easily through a random process, we have no sufficient statistical reason to consider it capable of generating profits in the future.

## Summary of the 4 fundamental steps of the Binomial evolution function

### Step 1: Transformation of the Trade Series

**Purpose**: To test whether two consecutive forecasts of the same type (two *buys* or two *sells*) are truly independent.

**Method**:

- The trade sequence is transformed by always alternating *buy* and *sell*.
- If there are two consecutive *buys* or *sells*, an opposite “test” operation is inserted between the two, of the same duration as the waiting time between them (ΔT).

**Example:**

Original sequence:

Buy — ΔT1 — Buy — ΔT2 — Buy — ΔT3 — Buy

Transformed sequence:

Buy — Sell(ΔT1) — Buy — Sell(ΔT2) — Buy — Sell(ΔT3) — Buy

**Step 2: Convert to binary sequence**

**Purpose**: To prevent a single large gain (or loss) from influencing the statistical analysis.

**Method**:

- Each trade is converted to:
  - 1 if it closed in profit
  - 0 if it closed in loss

**Example** (results of each trade in the transformed sequence):

Buy (profit) → 1

Sell (loss) → 0

Buy (profit) → 1

Sell (loss) → 0

Buy (profit) → 1

Sell (profit) → 1

Buy (loss) → 0

**Binary result:**
[1, 0, 1, 0, 1, 1, 0]

**Step 3: Selection of Independent Trades**

**Purpose:** Select trades that have been executed on different market evolutions.

**Method**:

- In a binary sequence, only consecutive trade pairs having the same results are accepted (1 after 1 or 0 after 0).
- This is because, to have the same result with two opposite operations (buy/sell), the system

must be changed ⇒ independence.

- Consecutive trade pairs where the results alternate (1 after 0 or 0 after 1) are discarded, because they are compatible with a stationary phase ⇒ possible dependence.

**Example** (using the sequence [1, 0, 1, 0, 1, 1, 0]):

Checks:

- 1 → 0 ⇒ rejected [**1, 0**, 1, 0, 1, 1, 0]
- 0 → 1 ⇒ rejected [1, **0, 1**, 0, 1, 1, 0]
- 1 → 0 ⇒ rejected [1, 0, **1, 0**, 1, 1, 0]
- 0 → 1 ⇒ rejected [1, 0, 1, **0, 1**, 1, 0]
- 1 → 1 ⇒ included [1, 0, 1, 0, **1, 1**, 0]
- 1 → 0 ⇒ rejected [1, 0, 1, 0, 1, **1, 0**]

Independent trades: only the pair 1 → 1 remain which corresponds to **two consecutive profitable trades**.

**Step 4: Randomness test using the Binomial distribution**

**Purpose**: To assess whether the observed results can be attributed to a random process, for example a fair coin toss.

**Method**:

- We consider:
  - k = number of successes (2)
  - n = number of independent trades selected (2)
  - p = 0.5 (fair coin)
- We calculate $P(X \geq k)$ with the binomial distribution.

**Example**:

- We obtained 2 successes on 2 independent trades.
- Using the binomial distribution: $P(X \geq 2) = 25\%$.
- The value is high ⇒ the result is compatible with a random process (there is no evidence of deterministic prediction).

If the observed number of successes is highly **improbable** under the assumption of randomness (probability < 1%), this provides statistical evidence that the prediction algorithm may exhibit **deterministic behavior**. Conversely, high probability values indicate that the observed results may plausibly be the outcome of a random process.

## How to distinguish a real "fortune teller" from simple luck using the Bionomial evolution function in the paradox of professional Traders

In the article, of the professional trader paradox [4], I develop a simple experiment based on the toss of a coin to explain the importance of considering financial markets as non-stationary. In this paragraph, I explain how the Binomial evolution function can correctly analyze the results of this experiment. In this way, we could understand if the strategy used has useful information that allows it to increase the probability of prediction.

### 1. The Experiment

Two people, **A** and **B**, sit opposite each other, separated by a black cloth.

- **A** can toss a coin as many times as he wants during a fixed interval of time.
- **B** does not see the coin but, at any time, can "bet" by declaring whether the coin at that moment shows **Heads** or **Tails**.
- If the declaration coincides with the real state of the coin, **B** obtains a "success".

The question: **does B really know how to "read" the coin or does he win only by chance?**

### 2. Why the *Binomial evolution function* is needed

In this experiment, **A** flips freely without being seen and **B** tries to guess Heads/Tails at any time. From the outside, **we do not see the flips so we have the following problem**:

- If **B** hits **two or more identical predictions in a row,** we can interpret it in two opposite ways:
    1. **B is really skilled** and anticipates every single flip.
    2. **B is just "breaking up" the same bet**: the coin is not flipped again and he repeats the same answer on a single event already decided.

Confusing the two cases means mistaking a mere accounting trick for a "winning strategy." Understanding *when* we are truly predicting new coin tosses and *when* we are simply spreading our bets over the same toss is therefore, crucial for evaluating any betting method.

### 3. Where the Binomial evolution function intervenes

The *Binomial evolution function* (BEF) was invented precisely to separate these two situations, here's how:

| | |
|---|---|
| **Transformation** | It imposes the alternation *Heads-Tails* between two successive predictions, inserting a "dummy toss" if necessary. |
| **Binary conversion** | Each prediction becomes 1 (correct) or 0 (wrong); the individual amounts no longer matter, so each win and each loss have the same statistical value. |
| **Selection of independent cases** | Compares the result of each prediction with the one immediately before: it only keeps 0-0 or 1-1. Indeed, **two equal and opposite predictions give the same result, only if there has been a new toss**. |
| **Binomial test** | Calculates the probability that the independent successes of **B** are the result of a random process. |

**3. What we discover**

- **If the p-value remains large**, the consecutive wins are nothing special: they are consistent with a simple "splitting" of a single bet over one toss.
- **If the p-value is low**, then it means that the consecutive victories refer to effectively different throws and B has real information on the individual events.

When betting on non-stationary systems, one must analyze very carefully streaks of victories that derive from forecasts all in the same direction. In finance, the formation of groups of high returns, alternating with groups of low returns, creates a statistical phenomenon called clustering [5]. In other words, it means that the returns are not distributed homogeneously but tend to group together. The clustering phenomenon can have significative deleterious effects in finance, as it leads to the misperception that trades executed during winning streaks are statistically independent, when in fact they may be correlated. Consequently, it becomes essential to understand if the predictions were **on truly distinct events** by applying the Binomial evolution function to filter the self-delusion of "splitting the stakes" on the same roll.

Only in this way, do we know if we are acting as intelligent strategists or if we are simply multiplying bets on the same roll, convincing ourselves that we are winning when, in reality, we are acting irrationally.

## Conclusion

The Binomial Evolution Function represents a new class of functions that address the problem of non-independence of results in finance. Resolving this problem has become a central challenge in the modern approach to quantitative trading. As a result, there is growing interest in methods that can transform sequences of outcomes into statistically independent data. The method used by the Binomial evolution function to select independent data is conceptually simple: it is based on the idea that **two identical but opposite predictions yield the same result only if the underlying system has evolved**. This insight provides a practical criterion for identifying independent trades.

In conclusion, this function allows us to answer the following fundamental question:

*Am I betting on different events or am I simply splitting a bet on the same event?*

If we cannot answer this question, any analysis of the results will be useless.

## References:

[6] Fama, E. F. (1965). "The Behavior of Stock-Market Prices." The Journal of Business, 38(1), 34-105.

[7] O. Peters, M. Gell-Mann. "Evaluating gambles using dynamics", Chaos: An Interdisciplinary Journal of Nonlinear Science, 2016; 26 (2): 023103 DOI: 10.1063/1.4940236.

[8] Berdondini, Andrea, Application of the Von Mises' Axiom of Randomness on the Forecasts Concerning the Dynamics of a Non-Stationary System Described by a Numerical Sequence (January 21, 2019). Available at SSRN: https://ssrn.com/abstract=3319864 or http://dx.doi.org/10.2139/ssrn.3319864.

[9] Andrea Berdondini, "The Professional Trader's Paradox", (November 20, 2018). Available at SSRN: https://ssrn.com/abstract=3287968.

[10] Marti, G., Nielsen, F., Bińkowski, M., & Donnat, P. (2017) "A review of two decades of correlations, hierarchies, networks and clustering in financial markets", arXiv preprint arXiv:1703.00485.

# Psychology

The first article deals with the mental predisposition called "dissociation from the result" that characterizes professional traders. This attitude is fundamental, because in the financial markets, being characterized by a low number of degrees of freedom, it is particularly easy to obtain good results randomly. Therefore, in these conditions, the result can be a little significant. Consequently, developing a dissociation from the result is essential in order not to overestimate your trading strategy. Indeed, future results depend on our knowledge of the system on which we make predictions and not on past results.

In the second article, we show how meditation is an efficacious mental training method to improve our approach to problem solving. The importance of this type of practice derives from its ability to reduce our irrational and therefore random actions. As we have seen in previous articles, acting randomly is not only useless but also increases the uncertainty of future results. Therefore, this last article, despite talking about a topic that seems to be distant from the topic covered in this book, completes all the concepts discussed previously.

# The psychology of the professional trader

Andrea Berdondini

ABSTRACT: In this article, we will analyze a mental attitude that distinguishes professional traders. This characteristic can be summarized with the following concept: *"dissociation from the result"*. This type of mental predisposition is very important, because the way we relate to the result affects our ability to behave rationally.

## The basic feature of the human mind

To understand how the mind behaves in relation to the result of the actions, we must understand the environment in which it evolved. Man has evolved in a context in which the link between action and result has always been very strong.

The non-randomness of the result has an important consequence: if the action leads to a benefit, the action that generated it turns out to be correct. Therefore, a very strong connection is created between the action and the result. Consequently, to a useful result for our survival we are led to consider the action performed as rational and correct. The brain also strengthens this bond in a physiological way by producing a physical sensation of happiness. In practice, the brain rewards us for the action done and wants it to be repeated over time because considered useful for our survival.

## How the mind is deceived

Now we analyze how a mind, evolved under the circumstances described in the previous paragraph, behaves when it has to face situations in which the outcome of the result is subject to a significant random component. In these situations, an extremely important thing happens: the result of our action will never be constant but will have a certain degree of randomness.

In this situation, the link between action and result is broken, and it is no longer true that a useful result corresponds to a rational action. Situations in which there is no link between the rationality of the action and the result are situations almost exclusively created by man as in the case of gambling. For example, if we play heads or tails, we have 50% of probability to win and 50% to lose, and obviously, there is no rational link between action and result. However, for the brain it makes no difference, if you win a bet it will reward you with a feeling of happiness and this is because for our mind the link between result and action still exists. For this reason, the psychological pathology called ludopathy, that afflicts gamblers, is the direct consequence of an evolved mind in an environment where there is a very strong link between action and result.

In this situation, the mind is deceived and considers actions that are irrational and destructive as rational and useful. This behavior is also called cognitive distortion, and the people who are affected are really convinced that their irrational actions can lead them to victory.

At this point, you will surely have understood that the main problem is due to the direct link that the mind associates between action and result. Therefore, it is precisely on this key element that arises the difference in approach between an ordinary person and a professional trader.

People who understand this problem understand the importance of training their mind about the dissociation from the result. In this way, they develop a new awareness about the result, which allows them to maintain a rational and winning response even in situations where the result is subject to a random component.

## The mind of a professional trader

In the previous chapter, the cognitive characteristic that distinguishes an ordinary person from a professional trader was identified. This characteristic can be summarized with the following concept: "dissociation from the result". In other words, it is about breaking the link that our brain creates between action and result.

This mental approach is present among professionals from very different sectors; for example, it is normal for a professional poker player to get angry when he makes a stupid play and to be happy when he makes an intelligent play, regardless of the outcome of the individual plays. In fact, these players have perfectly understood that there is no direct link between the result of a single operation and the action performed. What makes them successful is always being able to perform a series of rational and correct actions without getting involved in emotions. About this argument, I also report an interesting statement by the professional trader Linda Raschke whose interview was featured in the 06-2017 issue of Trader Magazine. When asked, "*Is Trading a game for you?"* Linda Raschke replies *"Yes. I barely check the account balance. Because this unnecessarily affect me both positively and negatively and does not change the fact that I have to make the right decision again. And if you always make the right decisions, performance becomes simple in the long run"*. As you can see, there is such a strong dissociation from the single result that she does not even care about its balance (an amateur trader checks his balance every 5 seconds).

Viewed from this point of view, trading is no longer random, but is a purely cognitive process, where you only operate if you understand the financial markets by always performing rational and never emotional actions. At this point, becomes easy to understand the difference between an amateur trader and a professional trader: the first is obsessed by the result, while the second is obsessed by the knowledge.

# How meditation can improve our problem-solving skills

Andrea Berdondini

ABSTRACT: The goal of this article is to try to explain easily, how the practice of meditation affects our problem-solving skills. To do this, I use a mathematical logical method to characterize problems based on the number of their possible solutions. In this way, it easy to explain the two primary approaches that the mind uses in solving problems. The first approach is iterative, optimal in solving simple problems (problems with a low number of solutions). The second approach is the logical one, optimal in solving complex problems (problems with a high number of solutions). The interesting aspect of these two methods is that their mode of action is the opposite. The iterative method is based on action, while the logical method is a reflective approach, in which any unnecessary action takes us away from solving the problem. Consequently, we will see how through the practice of meditation, we can shift our mental predisposition towards the logical approach by inhibiting our propensity for the iterative approach. This is a very important result because it can allow us to improve our ability to solve complex problems. This type of attitude is fundamental in a society where technological progress is making all simple and repetitive jobs less indispensable.

## The two fundamental methods used by the mind in solving problems

From a mathematical logical point of view, the problems can be divided according to the number of possible solutions they can have. Therefore, using this approach, we can define two classes of problems:

1) “Simple” problems: the problems in which the space of possible solutions is constituted by a small number of elements. Where the time required to try all possible solutions iteratively, is limited and acceptable. Example: a padlock that has 100 possible combinations, if I can try a different combination every 5 seconds, I will try all the combinations in an acceptable time.

2) “Complex” problems: the problems in which the space of possible solutions is constituted by a large number of elements. Where the time to try all possible solutions iteratively, tends to infinity or to an unacceptable time. Example: proving a mathematical theorem, doing it iteratively trying random solutions, takes a time that tends to infinity.

Thanks to this classification, we can study, in a simple way, the two fundamental approaches that the mind uses in solving problems.

The first approach is iterative, the mind does not try to solve the problem but tries all possible combinations. This approach is the optimal one for solving “simple” problems and was fundamental in the initial part of human evolution.

The second approach is the logical one; the mind creates a model of the problem and tries to solve it. This approach is the optimal one for solving “complex” problems, and its importance has increased in the course of human evolution. Consequently, this category of problems is also the one that most characterizes our problem-solving skills.

Comparing these two methods, the interesting thing we notice is that they act oppositely. The iterative method is based on action, faster I act, faster I solve the problem. Instead, the logic-based method is thoughtful; any incorrect action takes us away from the solution.

To explain the importance of not acting irrationally, when trying to solve a complex problem, I like to give the following example: imagine that you are a hiker who got lost in the jungle, what are you going to do? If we try to ask this question to a survival expert, he will answer that the best thing to do is to do nothing, and wait for help, because any of your actions will only tire you and put you in danger. The same thing happens in solving complex problems, in which every irrational action is not only useless but it makes us lose energy and time. This example makes us understand how different the two mental approaches are, and how fundamental our mental predisposition is to be successful in situations where a type of problem predominates.

Another useful point of view, to understand the importance of these two mental approaches, is to comprehend why there is so much interest in algorithms based on artificial intelligence. The reason for such interest stems from the fact that through artificial intelligence the algorithms are moving from an iterative approach to a logical approach. In fact, for example, the software developed to play chess, until recently used iterative approaches. In practice, the software simulates all possible combinations and chose the best move. This method had two important limitations: it needed a very powerful computer and could not be applied to games like the "go" in which the possible move combinations are very high. With the advent of artificial intelligence, these virtual players have gone from an iterative approach to a logical approach with incredible results. Google's Deepmind research team has developed the first software capable of beating the human champion of "go", on this topic I recommend reading the article published on nature "Mastering the Game of Go without Human Knowledge".

Now you can understand why the knowledge of these two different mental approaches is fundamental for studying the dynamics that involve our problem-solving skills.

## Meditation as mental training to improve the problem-solving skills

In this section, we will try to explain the implications of meditation on problem-solving skills. The term meditation refers to a large number of techniques, even very different from each other, whose task is to bring complete awareness of the present moment. One of the oldest and best-known techniques, and consequently among the most practiced, is called vipassana. The practice of this meditation is performed by sitting cross-legged while remaining completely still in a mental state in which we observe everything that happens. Mainly the observation is directed towards thoughts that tend to manifest themselves and towards one's breathing.

If we now analyze the two mental approaches, described in the previous paragraph, it is easy to understand how the practice of this type of meditation tends to be in contrast with the iterative method used in solving problems. As described in the previous paragraph, this approach is based on action, in practice, I act as quickly as possible without ever stopping. Hence, sitting still for no purpose represents the opposite of this method.

Consequently, the constant and repeated practice of this type of meditation leads over time to inhibit our propensity to act impulsively. There are many scientific studies on this topic that show how meditation reduces our propensity to multitasking (hyperactivity) and all those irrational and emotional behaviors. This is an important fact because the iterative approach is based on random (irrational) and continuous actions with a strong emotional component.

Meditation in this way modifies our problem-solving skills making us more reflective,

consequently increasing our propensity towards the use of the logical approach in solving problems. This result is significant because when we talk about problem-solving we are talking, in most cases, about the ability to solve complex problems. In fact, in a society where technological progress has an exponential trend, our ability to solve problems of this type becomes an increasingly important and requested skill.

Another fundamental aspect to keep in mind, regarding the importance of training the mind to a more reflective approach, is to understand the impact that new technologies have on our minds. To answer this question, we need to understand how most of the applications that are used on smartphones, tablets, etc. are developed.

The main purpose of these applications is to create an addiction, and to do this they take advantage of the iterative approach that the mind uses to solve simple problems. This is done because, in this situation, the person is forced to perform a continuous series of actions, which will correspond to a series of results, the consequence of which is a stimulation of the reward system present in our brain. With this technique, the user of the application will compulsively experience a succession of emotions, the result of which is to create a real addiction.

In conclusion, this type of technology is changing the approach to solving the problems of the new generations, favouring the iterative approach over the logical one. For this reason, it is essential to counteract the conditioning caused by these applications with techniques such as meditation, which inhibit our propensity to solve problems iteratively.

## Conclusion

In this article, I have used a simple mathematical logical analysis to relate our problem-solving skills and the practice of meditation. In this way, we find a similarity between two very different realities. On the one hand, we have a scientific formalism, in which through the analysis of a mathematical data the optimal approach is found to solve a class of problems. On the other hand, we have meditation, which represents a topic mainly studied in the philosophical field. So we have two extremely distant points of view which, however, as we have seen, tend to have incredibly similar elements of convergence. Indeed, the practice of meditation represents a way of acting contrary to the iterative approach. Consequently, meditation acts by inhibiting our propensity to act iteratively, making us prefer the logical approach, fundamental in solving complex problems. We have also seen how modern technologies are influencing new generations to hyperactive and compulsive approaches. So, it becomes essential to contrast this type of mental conditioning, with something that goes to act in the opposite direction leading us to act more thoughtful. Meditation, from this point of view, can be seen as a practice that acts on some primary aspects used by the mind in many of its processes, such as problem-solving. In this way, we can partly understand why something so simple has such profound implications in many areas of the brain. For these reasons, I believe that meditation will become an increasingly important formative practice.

# Risk Management (Good News)

In this final chapter, we talk about risk management, one of the topics that most interests the investor. In addition, this chapter summarizes all the concepts discussed above.

As you may have noticed, I have inserted the phrase "good news" in the background, the reason for this sentence is the following: in a world like that of trading where it is difficult to have certainties, risk management is defined by some irrefutable rules. For this reason, it represents a solved problem.

The action of investing can be divided into two parts:

1 the choice or development of an investment method.

2 The allocation of capital to be associated with the chosen investment method.

The first fundamental rule of risk management is as follows:

*“Risk management concerns solely the choice of the ratio between invested capital and liquidity”*

So, it's just about point 2. As a result, it only comes into play when we need to decide how much money to invest on the chosen investment method.

The second fundamental rule of risk management is as follows:

*"Risk management should not influence in any way the choice or development of an investment method”.*

The most common mistake made in finance is to introduce risk management into the choice of an investment method. In this way, the emotional component, which is part of risk management, alters our ability to rationally choose the best investment method.

A similar behaviour happens in poker, also in this case, we must choose a game strategy and a capital to invest. The mistake that the amateur poker player makes is the following: the player is afraid of losing money, consequently he plays in a very closed way (never bluffing) and therefore his plays are easily predictable. Furthermore, he plays at tables with very high stakes. In this way, the player performs risk management by changing the game strategy and not on the invested capital (stakes with which he sits at the table).

The investor makes the same mistake as the amateur poker player, by implementing risk management in choosing the investment strategy and not in choosing the capital to invest. As a result, he finds himself in a situation where he is investing too much money on a strategy that is not the best.

I conclude with a very simple example: let's take one hundred investors, all with different characteristics both from an emotional and economic point of view. These hundred investors can access the same investment strategies. What is the optimal situation for these investors? The optimal choice, for these investors, is that everyone should use the same strategy, the best strategy (less random), but they should invest different amounts based on their emotional and economic characteristics.

In summary, the choice of investment strategy is independent of the characteristics of the investor. Instead, the choice of the amount to invest depends on the characteristics of the investor and determines risk management.

# Sticker for your monitor

Cut out the sentence below and apply it on your work monitor.

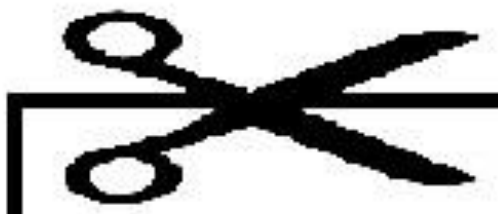

*"The only thing that cannot be created randomly is knowledge"*

Trying to create knowledge randomly is one of the most favourite hobbies for traders. This sentence is intended to remind you of the uselessness of this way of acting.